\documentclass[letterpaper,12pt]{article}
\usepackage{authblk}
\usepackage[utf8]{inputenc}
\usepackage[english]{babel}
\usepackage{amsmath}
\usepackage{amssymb}
\usepackage{graphicx}
\usepackage{multirow}
\usepackage{epsfig}
\usepackage{rotating}
\usepackage{longtable} %\usepackage[english]{babel}
\usepackage{caption}
\usepackage{subcaption}
\usepackage{url}
\usepackage{pdflscape}
\usepackage{lscape}
\usepackage{cancel}
\usepackage{setspace}
\pdfoutput=1
\usepackage{csquotes}
\usepackage{placeins}
\usepackage[left=2cm,top=3cm,right=2cm,bottom=3cm]{geometry}
\usepackage{tabularx}
\usepackage{adjustbox}
\usepackage{color,xcolor}

\usepackage{hyperref}
\usepackage{comment}
\usepackage{mathtools}
\usepackage{latexsym}

\newcommand{\ie}{{\emph i.e.,\ }}

\title{The Standard Model of Particle Physics as an effective theory from two  non-universal $U(1)$'s}

\author[a]{Richard H. Benavides \footnote{richardbenavides@itm.edu.co, corresponding author.}
}
\author[b]{Yithsbey Giraldo \footnote{yithsbey@gmail.com}  }
\author[c]{William A. Ponce \footnote{william.ponce@udea.edu.co}}
\author[c,d]{Oscar Rodríguez \footnote{oscara.rodriguez@udea.edu.co}}
\author[b]{Eduardo Rojas. \footnote{eduro4000@gmail.com} }

\affil[a]{Instituto Tecnol\'{o}gico Metropolitano, Facultad de Ciencias Exactas y Aplicadas,  \textit{\small Calle 73 N° 76-354 vía el volador, Medell\'{i}n, Colombia.} }

\affil[b]{Departamento de Física, Universidad de Nariño, \textit{\small Calle 18 Carrera 50, A.A. 1175, San Juan de Pasto, Colombia}}

\affil[c]{Instituto de Física, Universidad de Antioquia,  \textit{\small A.A. 1226, Medellín, Colombia.}}

\affil[d]{Facultad de Ingenierías, Universidad de San Buenaventura, \textit{\small Carrera 56C N° 51-110, Medellín, Colombia. }}

\begin{document}
  
\maketitle

\abstract{We study the possibility of obtaining the Standard Model (SM) of particle physics as an effective theory of a more fundamental 
one, whose electroweak sector includes two non-universal local $U(1)$ gauge groups, with the chiral anomaly cancellation taking place through an interplay among families.  As a result of the spontaneous symmetry breaking, a massive gauge boson $Z'$ arises, which couples differently to the third family of fermions (by assumption, we restrict ourselves to the scenario in which the $Z'$ couples in the same way to the first two families). Two Higgs doublets and one scalar singlet are necessary to generate the SM fermion masses and break the gauge symmetries.  
We show that in our model, the flavor-changing neutral currents (FCNC) of the Higgs sector are identically zero if each right-handed SM fermion is only coupled with a single  Higgs doublet. This result represents a FCNC cancellation mechanism different from the usual procedure in Two-Higgs Doublet Models~(2HDM).
The non-universal nature of our solutions requires the presence of three right-handed neutrino fields, one for each family. Our model generates all elements of the Dirac mass matrix for quarks and leptons, which is quite non-trivial for non-universal models. Thus, we can fit all the masses and mixing angles with two scalar doublets. Finally, we show the distribution of solutions for the scalar boson masses in our model by scanning well-motivated intervals for the model parameters. We consider two possibilities for the scalar potential and compare these results with the Higgs-like resonant signals recently reported by the ATLAS and CMS experiments at the LHC.
Finally, we also report collider, electroweak, and flavor constraints on the model parameters.} 

\section{Introduction}

The Standard Model of particle physics (SM) based on the local gauge group $SU(3)_\text{C}\otimes SU(2)_L\otimes U(1)_Y$ \cite{donoghue_golowich_holstein_2014} has been very successful so far, in the sense that its predictions are in good agreement with the present experimental results, including the latest discovery of the Higgs boson \cite{Nisati:2017cst, ParticleDataGroup:2022pth,ATLAS:2016neq}, a fundamental ingredient of the model that contributes to our understanding of the origin of mass for the subatomic particles. However, the SM fails short in explaining things as: hierarchical charged fermion masses, fermion mixing angles, charge quantization, strong CP violation, replication of families, neutrino masses and oscillations, and the matter-antimatter asymmetry of the universe. Besides, gravity is excluded from the context of the model and good candidates for dark matter and dark energy present in the universe are not provided~\cite{Ma:1990ce, Langacker:1980js, Dine:1981rt, Benavides:2022hca, Alonso:2013nca, Mohapatra:2006gs, Perlmutter:1999rr,Joyce:2014kja}.

Replication of families, also known as the ``family problem'', refers to the fact that the SM is not able to predict the number $N$ of fermion families existing in nature, something related with the universality of the model, which means that the gauge anomalies, in particular those associated with the $U(1)_Y$ hypercharge, cancel out exactly for each family; the only restriction, $N\leq 8$, comes from the asymptotic freedom of $SU(3)_\text{C}$ also known as quantum chromodynamics or QCD~\cite{Golowich:1990ki}. Experimental results at the CERN-LEP facilities early in the 1990s implied the existence of at least three families, each one having a neutral lepton with a mass less than half the mass of the neutral $Z$ gauge boson \cite{ALEPH:1989kcj}; this result was initially interpreted as an exact value for the total number of families in nature, which is not quite correct. As a matter of fact, the LEP data does not exclude the existence of additional families having heavy neutrinos.

Therefore, it is widely believed that the SM is not truly fundamental, with the prevailing view that the model is just a low-energy effective description of a more complete theory. There are several good candidates for this, all of them grouped in what is now known as ``the physics beyond the Standard Model'' (BSM)~\cite{Ellis:2012zz,Capozziello:2011et,Abdalla:2022yfr}. {Thus, there are numerous works with gauge extensions of the $U(1)$ type, either to explain neutrino masses or dark matter, etc., see references:\cite{Davidson:1979wr,Marshak:1979fm,Ma:2001kg,Barr:2005je,Adhikari:2008uc,Okada:2012sg} as to show some of them. However, our goal is to introduce two non-universal $U(1)$  symmetry gauge to SM to obtain it as an effective model.}
{
The consideration of \( Z' \) bosons with non-universal couplings is justified for theoretical and experimental reasons. From a theoretical perspective, these models  arise naturally in several scenarios, for example in string models\cite{Langacker:2000ju,Langacker:2008yv,CrispimRomao:2017tnd,Antusch:2017tud}
 and 3-3-1~models\cite{Pisano:1992bxx,Frampton:1992wt,Pleitez:2021abk,Benavides:2021pqx,Suarez:2023ozu}.
 However, from a phenomenological point of view, 
they are  convenient for studying experimental anomalies at low energies, for example: anomalous decays of $B$-mesons~\cite{Benavides:2016utf,Tang:2017gkz,Maji:2018gvz,Garcia-Duque:2021qmg,Athron:2023hmz}, Cabibbo angle anomaly~\cite{Alok:2020jod}, muon anomalous magnetic moment(or muon $g\text{-}2$)~\cite{Alvarado:2021nxy,Frank:2021nkq,Greljo:2022dwn} and rare charm decays~\cite{Alok:2021pdh}.
Recently CMS reported for the first time searches for neutral vector bosons with non-universal couplings~\cite{CMS:2024xuo} due to the multiple applications of this class of models.
Thus, searching for signals associated with these models remains a relevant task in exploring physics beyond the Standard Model~\cite{Salvioni:2009jp}.
}

In what follows, and in order to shed some light on the shortcomings of the SM, we propose an extension of it; that is, a new model for three families based on the local gauge group $SU(3)_\text{C}\otimes SU(2)_L \otimes U(1)_\alpha \otimes U(1)_\beta$, where the charges associated with the two Abelian factors are non-universal, in the sense that they are not the same for the three assumed families. The fermion content of our model is the same as that of the SM, extended with three right-handed neutrinos  $\nu_{iR}$ ($i=1,2,3$), one for each family. 

\section{The model}

In this section, we elaborate on the mathematical aspects of the new model in consideration, which is a minimal extension of the SM, both in its gauge sector and in its fermion sector. As a consequence, the scalar sector must also be enlarged, something we are going to do in the most economical possible way.

As mentioned above, the model to be considered here is based on the local gauge symmetry $SU(3)_\text{C}\otimes SU(2)_L \otimes U(1)_{\alpha} \otimes U(1)_{\beta}$, where $SU(3)_\text{C}$ and $SU(2)_L$ are the same as in the SM, and the two Abelian factors are non-universal, capable of projecting the SM $U(1)_Y$ hypercharge to a lower energy scale. So, as a result of the spontaneous symmetry breaking, a new gauge boson associated with a non-universal neutral weak current appears.

The fermion fields in our model are the same as in the SM, together with three neutral Weyl states associated with the three right-handed neutrino components, one for each family. This popular fermion extension of the SM has been used to explain neutrino masses and oscillations, the baryon asymmetry of the Universe, dark matter and dark radiation, and in our approach, it has the peculiarity  that, unlike what happens in the SM, the three new fields have non-vanishing charges under both $U(1)$ factors.

As for the scalar sector, we first  introduce a SM singlet field $\sigma$ able to spontaneously break the $U(1)_{\alpha} \otimes U(1)_{\beta}$  symmetry  down to $U(1)_Y$. To break the remaining symmetry and at the same time implement the Higgs mechanism, at least one $SU(2)_L$ scalar doublet $\Phi_2$ (developing a vacuum expectation value (VEV) at an energy scale $v_2$) must be introduced, in such a way that the remaining symmetry $SU(3)_\text{C}\otimes U(1)_Q$ survives down to laboratory energies. We choose the quantum numbers of this doublet such that it only provides  tree-level masses to the third fermion family. To generate (at tree level) the other fermion masses and the mixing matrices, at least one more $SU(2)_L$ scalar doublet $\Phi_1$ must be included. This doublet develops a vacuum expectation value~(VEV) at an energy scale $v_1 < v_2$.

Table \ref{notation} shows the fermion and scalar content of our model, along with the notation used for the different Abelian charges, as well as the weak-isospin $T_{3}$, hypercharge $Y$, and electric charge $Q$ of the particles. In our analysis, we will assume that  $\chi_{f_1}=\chi_{f_2}\neq \chi_{f_3}$, where $\chi_{f_i}$ stands for the Abelian $\alpha, \beta$ charges, $f=q,u,d,l,\nu,e$ and $i=1,2,3$, that is, we consider a model with universal couplings for the first two fermion families, but not for the third one, a convenient condition in the implementation of models with minimal flavor violation, that in turn provides a way to distinguish the third family from the first two ones. In this way, our model is characterized by 24 parameters associated with the fermion sector and 6 more with the scalar one, for a total of 30 free parameters which can be fixed by demanding a renormalizable model, reproducting the SM hypercharges, and appropriate Yukawa couplings to provide fermion masses. 

\subsection{Cancellation of chiral anomalies}

Regarding the renormalizability of the theory, we must ensure an anomaly-free scenario, which is achieved by imposing the following relations among the $U(1)$ fermion  charges: 
\begin{align}
[SU(3)_\text{C}]^2\otimes U(1)_\alpha:\ &\sum_i \left(2\alpha_{qi}-\alpha_{ui}-\alpha_{di}\right)=0,\notag\\
[SU(2)_L]^2\otimes U(1)_\alpha:\ &\sum_i \left(3\alpha_{qi}+\alpha_{li}\right)=0,\notag\\
[\text{grav}]^2\otimes U(1)_\alpha:\ &\sum_i \left(6\alpha_{qi}-3\alpha_{ui}-3\alpha_{di}+
2\alpha_{li}-\alpha_{\nu i}-\alpha_{ei}\right)=0,\notag\\
[U(1)_\alpha]^2 U(1)_\beta: \ &\sum_i \left(6\alpha_{qi}^2\beta_{qi}-
3\alpha_{ui}^2\beta_{ui}-3\alpha_{di}^2\beta_{di}+2\alpha_{li}^2\beta_{li}-
\alpha_{\nu i}^2\beta_{\nu i}-\alpha_{ei}^2\beta_{ei}\right)=0,\notag\\
[U(1)_\alpha]^3:\ &\sum_i \left(6\alpha_{qi}^3-3\alpha_{ui}^3-3\alpha_{di}^3+2\alpha_{li}^3-
\alpha_{\nu i}^3-\alpha_{ei}^3\right)=0,\label{anomalies}
\end{align}
together with the five corresponding equations  for the $U(1)_\beta$ group. These are obtained from the previous ones via the $\alpha \leftrightarrow \beta$ exchanging for a total of 10 equations. Given that the number of involved unknowns is greater (24 assuming universality in the first two fermion families), the number of possible solutions is infinite, so, just like in the SM, chiral anomaly cancellation is not sufficient to explain the charge quantization \cite{Golowich:1990ki}. 

\begin{table}[]
\begin{center}
\begin{spacing}{1.5}
\begin{tabular}{|c|c|c|c|c|c|c|c|c|}
\hline
 & $\ell_i\equiv(\nu_{iL},e_{iL})^T$  & $\nu_{iR}$  & $e_{iR}$  & 
$q_i\equiv(u_{iL},d_{iL})^T$ & $u_{iR}$ & $d_{iR}$ 
& $\Phi_a = (\phi^+_a,\phi^0_a)^T$ & $\sigma$  
\\ \hline
$\hat{T}_{3}$ & $(\frac{1}{2},-\frac{1}{2})^T$ & $0$ & $0$ 
& $(\frac{1}{2},-\frac{1}{2})^T$ & $0$ & $0$ & $(\frac{1}{2},-\frac{1}{2})^T$  
&$0$\\ \hline
$\hat{Y}$ & $-1$ & $0$ & $-2$ & $\frac{1}{3}$ & $\frac{4}{3}$ & $-\frac{2}{3}$ 
& $1$ & $0$  \\ \hline
$\hat{Q}$ & $(0,-1)^T$ & $0$ & $-1$ & $(\frac{2}{3},-\frac{1}{3})^T$ & $\frac{2}{3}$ 
& $-\frac{1}{3}$ & $(1,0)^T$ & $0$  \\ \hline
$\hat{\alpha}$ & $\alpha_{li}$  & $\alpha_{\nu_i}$  & $\alpha_{ei}$ & $\alpha_{qi}$  
& $\alpha_{ui}$  & $\alpha_{di}$ & $\alpha_{a}$ &$\alpha_\sigma$  \\ \hline
$\hat{\beta}$ & $\beta_{li}$  & $\beta_{\nu i}$  & $\beta_{ei}$ & $\beta_{qi}$  
& $\beta_{ui}$  & $\beta_{di}$  & $\beta_{a}$ &$\beta_\sigma$\\ \hline
\end{tabular}
\end{spacing}
\caption{Here, $i$ runs over the number of families $(i=1,2,3)$, and $a=1,2$.}
\label{notation}
\end{center}
\end{table}

\subsection{The Lagrangian of the model %Model Lagrangian}}
}
In our model, the covariant derivative $D_\mu$ for the electroweak (EW) sector is given by
\begin{align}
D^\mu = \partial^u +ig_L A^\mu_{j}\hat{T}_j +i\frac{g_\alpha}{2}B^\mu_\alpha \hat{\alpha}  + 
i\frac{g_\beta}{2} B^\mu_\beta \hat{\beta}\ ,\label{Du}
\end{align} 
where $\hat{T}_j$, $A^\mu_{j}$ (with $j = 1,2,3$) and $g_L$ denote, respectively, the generators, the gauge fields, and the coupling constant associated with the weak isospin gauge group $SU(2)_L$, while $\hat{\chi}$, $B^\mu_\chi$ and $g_\chi$, with $\chi=\alpha,\beta$, are the corresponding quantities related with the two Abelian $U(1)$ factors. The terms in the Lagrangian describing the relevant interactions in our analysis are then:
\begin{align}
\mathcal{L} \supset\  
&-V(\Phi_1,\Phi_2,\sigma) \notag\\
&+|D^\mu \Phi_1|^2 + |D^\mu \Phi_2|^2 + |D^\mu \sigma|^2 \notag\\
&+ i{\bar q_{j}}\cancel{D} q_{j} +i{\bar u_{jR}}\cancel{D} u_{jR}+ i{\bar d_{jR}}\cancel{D} d_{jR}+ i{\bar l_{j}}\cancel{D} l_{j} +i{\bar\nu_{jR}}\cancel{D} \nu_{jR}+ i{\bar e_{jR}}\cancel{D} e_{jR}\notag \\
 &-Y^e_{jk}\bar{\ell}_{j}\Phi_1 e_{kR}- Y^\nu_{ jk}\bar{\ell}_{j}\tilde\Phi_1\nu_{kR}- Y^d_{jk}\bar{q}_{j}\Phi_1 d_{kR}
 - Y^u_{jk}\bar{q}_{j}\tilde\Phi_1 u_{kR}\notag\\
&-Y^e_{j3}\bar \ell_{j}\Phi_2 e_{3R}- Y^\nu_{j3} \bar \ell_{j}\tilde\Phi_2\nu_{3R}-Y^d_{j3} \bar q_{j}\Phi_2 d_{3R}
- Y^u_{j3} \bar q_{j}\tilde\Phi_2 u_{3R}+ \text{h.c.}\ ,\label{eq:lagrangian}
\end{align}
where sum over repeated indices is implied, with $j$ and $k$ taking the values $\lbrace 1,2,3\rbrace$ and $\lbrace 1,2\rbrace$, respectively. The term in the first line denotes the scalar potential. 
Due to the non-universal character of our model, a single scalar doublet $\Phi_1$ is not enough to provide masses to all fermion particles and, simultaneously, to generate realistic mixing matrices. To this end, at least another Higgs doublet $\Phi_2$ developing a VEV is required. Additionally, a scalar singlet must be introduced to break the abelian symmetries.
{
The symmetry group $SU(2)\otimes U(1)\otimes U(1)$ has five generators, four of which are broken, such that at low energies only the electromagnetic gauge group $U(1)_{\text{QED}}$ survives.  For a model with just two Higgs doublets,  by applying the Higgs mechanism we obtain,  in addition to the SM fields,  two exotic fields:  a CP even neutral scalar field and a charged one, the remaining ones are absorbed as goldstone bosons by the vector fields. To accommodate the experimental anomalies it is necessary to include a scalar singlet to break the $U(1)\otimes U(1)$  at high energies such that, in addition to the SM fields we get two CP even scalar fields and a pseudoscalar. This is also convenient if we want the $Z'$ scale to be larger than the electroweak scale since the quadratic sum of the VEVs of the doublets must equal $v_{SM}= 246.24$GeV.
}
The scalar potential is analyzed in Appendix \ref{sec:potential}.  The terms in the second line correspond to the scalar-gauge interactions responsible for the masses and mixings in the gauge sector (see Appendix \ref{sec:gauge}). Terms in the third line give rise to fermion-gauge interactions, as discussed in Sec.~\ref{sec:EW-currents}, and the Yukawa couplings present in the model are shown in the fourth and fifth lines. The invariance of the Yukawa interaction terms  under the $U(1)_\alpha \otimes U(1)_\beta$ gauge symmetry  implies   the following relations between the $\chi(=\alpha,\beta)$ charges:
\begin{align}
    \chi_{lj}-\chi_{1}-\chi_{ea}&=0,\notag\\
    \chi_{lj}-\chi_{2}-\chi_{e3}&=0,\notag\\
    \chi_{lj}+\chi_{1}-\chi_{\nu a}&=0,\notag\\
    \chi_{lj}+\chi_{2}-\chi_{\nu 3}&=0,\notag\\
    \chi_{qj}-\chi_{1}-\chi_{da}&=0,\notag\\
    \chi_{qj}-\chi_{2}-\chi_{d3}&=0,\notag\\
    \chi_{qj}+\chi_{1}-\chi_{ua}&=0, \notag\\    
    \chi_{qj}+\chi_{2}-\chi_{u3}&=0. \label{eq:yukawas}
\end{align}

\subsection{Spontaneous symmetry breaking}

Our aim is to break the gauge symmetry of the model in two steps, namely,
\begin{align}
SU(2)_L \otimes U(1)_\alpha \otimes U(1)_\beta 
\stackrel{\langle\sigma\rangle}{\longrightarrow} 
SU(2)_L \otimes U(1)_Y \stackrel{\langle\Phi_a\rangle}{\longrightarrow}U(1)_Q\,
\end{align}

where $a={1,2}$. To achieve this, we allow the SM scalar singlet  $\sigma$ (charged under both $U(1)$'s factors) to acquire a VEV at a high energy scale, inducing a mixing between the $B_\chi$ fields that give rise to both: the SM gauge boson $B$ associated with the $U(1)_Y$ hypercharge symmetry and a new massive gauge boson $Z'$ with non-universal couplings to fermions. If $\theta$ is the angle parameterizing this mixing, then
\begin{align}
\begin{pmatrix}
B^\mu_\alpha\\
B^\mu_\beta
\end{pmatrix}=
\begin{pmatrix}
\cos \theta &-\sin \theta\\
\sin \theta &\cos \theta
\end{pmatrix}
\begin{pmatrix}
B^\mu\\
Z'^{\mu}
\end{pmatrix}.\label{firstBreaking}
\end{align}
Finally, at a lower energy scale (the EW one), the neutral components of the scalar doublets $\Phi_1$  and $\Phi_2$ develop  VEVs inducing the last breaking. Consequently, the $B$ and $A_{3}$ fields mix, giving rise to the massless photon $A^\mu$ and the massive SM neutral gauge boson $Z$. The corresponding mixing angle is the well-known Weinberg angle $\theta_W$:
\begin{align}
\begin{pmatrix}
A^\mu_{3}\\
B^{\mu}
\end{pmatrix}=
\begin{pmatrix}
\sin\theta_W &\cos\theta_W\\
\cos\theta_W &-\sin\theta_W
\end{pmatrix}
\begin{pmatrix}
A^\mu\\
Z^\mu
\end{pmatrix}.\label{secondBreaking}
\end{align}
The unbroken electric charge generator $\hat{Q}$ can be expressed as a linear combination of the three diagonal  (broken) generators of the gauge group after the spontaneous symmetry breaking, that is
\begin{align}
\hat{Q}=\hat{T}_{3L} +\frac{1}{2}\left(a_Y \hat{\alpha} +b_Y \hat{\beta}\right),
\end{align}
from which it follows that the SM hypercharge $\hat{Y}$ can be identified as 
\begin{align}
\hat{Y} = a_Y \,\hat{\alpha} +b_Y \hat{\beta},\label{Y}
\end{align} 
where $a_Y$ and $b_Y$ are two non-vanishing free parameters. However, these parameters turn be useless for our purposes, so they will be set to 1 for simplicity\footnote{From equation~\eqref{eq:YSM}, one of them can be absorbed in a redefinition of the scalar singlet hypercharges ($a_Y=-b_Y\beta_\sigma\alpha_\sigma$).}
In accordance with Eq. (\ref{Y}), the $U(1)$ charges displayed in Table \ref{notation} must satisfy the following relations:
\begin{align}
    &\alpha_{li}+\beta_{li}=-1,\notag\\
    &\alpha_{\nu i}+\beta_{\nu i}=0,\notag\\
    &\alpha_{ei}+\beta_{ei}=-2,\notag\\
    &\alpha_{qi}+\beta_{qi}=1/3,\notag\\
    &\alpha_{ui}+\beta_{ui}=4/3,\notag\\
    &\alpha_{di}+\beta_{di}=-2/3,\notag\\
    &\alpha_{a} +\beta_{a}=1,\notag\\
    &\alpha_{\sigma} +\beta_{\sigma}=0, \label{eq:YSM}
\end{align}
for $i=1,2,3$ and $a=1,2$. Thus, the breaking induced by the singlet $\sigma$ at an energy scale $v_\sigma$ allows to reproduce the SM hypercharges correctly.

\subsection{Mass and mixing matrices for fermions}
Let's now consider the generation of fermion mass, which takes place when $\Phi_2$ induces the breaking that gives rise to the local gauge $SU(3)_\text{C}\otimes U(1)_Q$ symmetry conserved at low energies. As mentioned, for non-universal models,  at least two scalar doublets are needed to provide masses to all the fermion particles and generate the mixing matrices.  As usual, the vacuum expectation values of the scalar doublets are given by 
\begin{align}
    \langle \Phi_a\rangle=\begin{pmatrix}
        0\\
        \dfrac{v_a}{\sqrt{2}}
    \end{pmatrix}\ ,\ \ (a=1,2)\ .
\end{align}
{In this model, it is possible to generate Dirac masses for all the standard model fermions including the SM neutrinos. In this case, the smallness of the neutrino masses relies on the Yukawa couplings as it does happen in the SM}.
The tree-level Dirac masses come from the Lagrangian (\ref{eq:lagrangian}). The resulting mass matrices take the form 
\begin{align}
M^f= 
\frac{1}{\sqrt{2}}\begin{pmatrix}
v_{1} Y^f_{11} &v_{1} Y^f_{12} &v_{2} Y^f_{13}\\
v_{1} Y^f_{21} &v_{1} Y^f_{22} &v_{2} Y^f_{23}\\
v_{1} Y^f_{31} &v_{1} Y^f_{32} &v_{2} Y^f_{33}
\end{pmatrix},
\end{align}
for $f=u,d,\nu,e$. From here, we see that despite the non-universality of the model, it is possible to have saturated mass matrices for leptons and quarks, \ie with all the matrix elements different from zero, which is a fairly non-trivial result. As a consequence, the CKM and PMNS mixing matrices can be easily generated, with the mixing between the first two fermion families induced by $\Phi_1$, while both $\Phi_1$ and $\Phi_2$ contribute to all the mixing elements involving the third family.\\

{Our model is capable of reproducing all the elements of the Dirac mass matrix (therefore, it has no texture zeros) so that it is always possible to reproduce the values of the masses and mixing angles, for both quarks and leptons~\cite{Branco:1999nb,
Gupta:2012fsl,Verma:2013qta,Sharma:2015gfa,Emmanuel-Costa:2016gdp,Benavides:2020pjx}.}

It is important to emphasize that all elements of the mass matrix are generated for each type of SM fermion, such that for each flavor there are 9 complex parameters,
{ which is equivalent to $2\times 18 = 36$ real parameters for both the up-type and the down-type quark mass matrices. Since the CKM matrix and the quark masses amount to 9 parameters and a phase, the number of free parameters exceeds the number of physical parameters to be fitted~\cite{Branco:1999nb}.
In the lepton sector, the neutrino masses are not known, and just two square mass differences are
experimentally available~\cite{ParticleDataGroup:2024cfk}; that is:
$\delta m_{21}=m^2_2-m^2_1$, $\delta m_{31}=m^2_3-m^2_1$ (where $\{ m_i\}_{i=1,2,3}$ are the neutrino masses), in this case there are only eight real parameters and a phase, but again the number of free parameters in the Yukawa couplings are enough to fit masses and mixing~\cite{Benavides:2020pjx}.
This freedom allows for the fitting of the quark and lepton masses and the CKM and PMNS mixing matrices with the most recent data from the literature~\cite{ParticleDataGroup:2024cfk}}.
%Al these results are in agreement with the results in the literature~\cite{Froggatt:1978nt,Branco:2000zi,Branco:1999nb}.

\subsection{Non-universal \texorpdfstring{$U(1)$}{} charges}

By solving the system of equations formed by Eqs. (\ref{anomalies}), (\ref{eq:yukawas}) and (\ref{eq:YSM}), we obtain a unique solution for the  $U(1)$ fermion and scalar charges. The resulting expressions, shown in Table \ref{tab:U1charges}, are given in terms of just three parameters, namely: $\alpha_{q1}$, $\alpha_{\nu 1}$ and $\alpha_{\nu 3}$\footnote{The $\alpha$ charge of the singlet $\sigma$, $\alpha_\sigma$, remains as a free parameter, but it does not affect the fermion charges, as can be seen in Table \ref{tab:U1charges}.}. From this it follows that the non-universality of the solution depends exclusively on the right-handed neutrino charges; so, in what follows, we will assume that $\alpha_{\nu 1}\neq \alpha_{\nu 3}$. Under this condition, the cancellation of chiral anomalies takes place among different families, and not family by family as it does in the SM. 

\begin{table}[]
\begin{center}
\begin{tabular}{|c|c||c|c||c|c|}
\hline
Field                       & $U(1)_\alpha$                    & Field      & $U(1)_\alpha$               & Field                     & $U(1)_\alpha$                                \\ \hline\hline
\multirow{2}{*}{$u_{iL}$}   & \multirow{4}{*}{$\alpha_{q1}$}   & $u_{aR}$   & $\alpha_{\nu 1}+4\alpha_{q1}$ & \multirow{3}{*}{$\Phi_1$} & \multirow{3}{*}{$\alpha_{\nu 1}+3\alpha_{q1}$} \\ \cline{3-4}
                            &                                  & $u_{3R}$   & $\alpha_{\nu 3}+4\alpha_{q1}$ &                           &                                              \\ \cline{1-1} \cline{3-4}
\multirow{2}{*}{$d_{iL}$}   &                                  & $d_{aR}$   & $-\alpha_{\nu 1}-2\alpha_{q1}$  &                           &                                              \\ \cline{3-6} 
                            &                                  & $d_{3R}$   & $-\alpha_{\nu 3}-2\alpha_{q1}$               & \multirow{3}{*}{$\Phi_2$} & \multirow{3}{*}{$\alpha_{\nu 3}+3\alpha_{q1}$}  \\ \cline{1-4}
\multirow{2}{*}{$\nu_{iL}$} & \multirow{4}{*}{$-3\alpha_{q1}$} & $\nu_{aR}$ & $\alpha_{\nu 1}$ &                           &                                              \\ \cline{3-4}
                            &                                  & $\nu_{3R}$ & $\alpha_{\nu 3}$ &                           &                                              \\ \cline{1-1} \cline{3-6} 
\multirow{2}{*}{$e_{iL}$}   &                                  & $e_{aR}$   & $-\alpha_{\nu 1}-6\alpha_{q1}$               & \multirow{2}{*}{$\sigma$} & \multirow{2}{*}{$\alpha_\sigma$}             \\ \cline{3-4}
                            &                                  & $e_{3R}$   & $-\alpha_{\nu 3}-6\alpha_{q1}$  &                           &                                              \\ \hline
\end{tabular}
\caption{Here $i=1,2,3$ and $a=1,2$. The corresponding $U(1)_\beta$ charges can be easily obtained by replacing $\beta$ instead of $\alpha$.}
\label{tab:U1charges}
\end{center}
\end{table}
{
As is well known from the literature on FCNCs, the
strongest constraints on tree-level flavor couplings come usually from $F^0-\bar{F}^{0}$ mixing processes $(F = K, B_d, D)$~\cite{Atwood:1996vj}, to avoid this problem with neutral scalar currents, 
in most models with two Higgs doublets, discrete symmetries are proposed to cancel the scalar currents with flavor changes. Four 2HDMs are known in the literature, Type-I, Type-II, Type-X, and Type-Y~\cite{Aoki:2009ha}, in these models each doublet is charged differently under the discrete symmetry such that one type of particle receives its mass from only one doublet. For example, type-up quarks receive the mass from the same scalar doublet. This is impossible for non-universal models since the charges $U(1)$ of the same type of quarks are different because the model is not universal. In our model, the up-type quarks of the first two families couple with one Higgs doublet while the up-type quark of the third family must couple with the other doublet. This result is guaranteed since the charges of the Higgs doublets are not equal. This is a different mechanism compared to the one in the models mentioned above; however, it can be considered as a particular case of condition III of the general theorem proved in  reference~\cite{Glashow:1976nt}.}

\section{EW currents and \texorpdfstring{$Z'$}{} couplings}\label{sec:EW-currents}

Ignoring the kinetic terms, the part of the Lagrangian (\ref{eq:lagrangian}) describing the interactions between fermions and gauge bosons can be written as
\begin{align}
-\mathcal{L}\supset \frac{g_L}{\sqrt{2}}\left(J^{\mu}_{W^+} W^+_\mu + \text{h.c.}\right)+
\frac{g_L}{2}J^\mu_{3}A_{3,\mu}+\frac{g_\alpha}{2}J^\mu_\alpha B_{\alpha,\mu}+\frac{g_\beta}{2}J^\mu_\beta B_{\beta,\mu}\ ,\label{L1}
\end{align}
where the $W^+_\mu$ field has been defined as $W^{+\mu}=(A^\mu_1-i A^\mu_2)/\sqrt{2}$ and the currents are given by
\begin{align}
J^{\mu}_{W}&={\bar\nu_{iL}}\gamma^\mu e_{iL} + {\bar u_{iL}}\gamma^\mu d_{iL}\ ,\notag\\
J^\mu_{3} &={\bar u_{iL}}\gamma^\mu u_{iL}+{\bar\nu_{iL}}\gamma^\mu \nu_{iL}-{\bar d_{iL}}\gamma^\mu d_{iL}-{\bar e_{iL}}\gamma^\mu e_{iL}\ ,\notag\\
J^\mu_\chi &=
\chi_{qi} \left({\bar u_{iL}}\gamma^\mu u_{iL}+{\bar d_{iL}}\gamma^\mu d_{iL}\right)+
\chi_{li} \left({\bar\nu_{iL}}\gamma^\mu \nu_{iL}+ {\bar e_{iL}}\gamma^\mu e_{iL}\right)\notag\\
&+\chi_{ui}\ {\bar u_{iR}}\gamma^\mu u_{iR}+\chi_{di}\ {\bar d_{iR}}\gamma^\mu d_{iR}+
\chi_{\nu i}\ {\bar\nu_{iR}}\gamma^\mu \nu_{iR}+ \chi_{ei}\ {\bar e_{iR}}\gamma^\mu e_{iR}\ , \label{Ja}
\end{align}
with a sum over the $i$ index is implied and $\chi=\alpha,\beta$. In the basis defined by (\ref{firstBreaking}), the Lagrangian in (\ref{L1}) can be expressed as
\begin{align}
-\mathcal{L}\supset \frac{g_L}{\sqrt{2}}\left(J^{\mu}_{W^+} W^+_\mu + \text{h.c.}\right)+\frac{g_L}{2}J^\mu_{3}W_{3\mu}+\frac{g_Y}{2}J^\mu_Y B_\mu +g_{Z'}J^\mu_{Z'}Z'_\mu\ ,\label{L2}
\end{align}
where the interactions of fermions with the $Z'$ boson are given by  
\begin{align}
g_{Z'}J^\mu_{Z'}&=g_\beta J^\mu_\beta\cos\theta -g_\alpha J^\mu_\alpha\sin\theta\notag\ ,\\
&=g_{Z'}\sum_{f}{\bar f_i}\gamma^\mu\left(\tilde\epsilon^L_{fi}P_L+\tilde\epsilon^R_{fi}P_R\right)f_i\notag \ ,\\
&=g_{Z'}\sum_{f}{\bar f_i}\gamma^\mu\left(\tilde{g}^V_{fi}-\tilde{g}^A_{fi}\gamma^5\right)f_i\ .
\end{align}
Here, $f$ runs over $\lbrace u,d,\nu,e\rbrace$, $i=1,2,3$ (corresponding with the SM family),  $P_{L,R}=(1\mp \gamma^5)/2$ are the chirality projectors and
\begin{align}
g_{Z'}\tilde\epsilon^{L,R}_{fi} =& \frac{1}{2}\left[g_\beta \hat{\beta}(f_{iL,R})\cos\theta-g_\alpha \hat{\alpha}(f_{iL,R})\sin\theta\right]\ , \label{gz'eLR}\\
\tilde{g}^{V,A}_{fi}=&\frac{1}{2}\left(\tilde{\epsilon}^L_{fi}\pm \tilde{\epsilon}^R_{fi}\right)\ .\label{gVA}
\end{align}
In Eq.~(\ref{gz'eLR}), $\tilde\epsilon^{L(R)}_{fi}$ denotes the left(right)-handed chiral coupling of the $f_i$ fermion to the $Z'$ boson, while in Eq.~(\ref{gVA}), $\tilde{g}^{V(A)}_{fi}$ represents the corresponding vector (axial-vector) coupling. As for the couplings to the $B$ field, we have that
\begin{align}
g_{Y}J^\mu_{Y}&=g_\alpha J^\mu_\alpha\cos\theta +g_\beta J^\mu_\beta\sin\theta\notag\ ,\\
&=g_{Y}\sum_{f}{\bar f_{iL}}\gamma^\mu \hat{Y}(f_{iL})f_{iL}+{\bar f_{iR}}\gamma^\mu \hat{Y}(f_{iR})f_{iR}\ ,
\end{align}
with 
\begin{align}
g_{Y}\hat{Y}(f_{iL,R}) = g_\alpha \hat{\alpha}(f_{iL,R})\cos\theta + g_\beta \hat{\beta}(f_{iL,R})\sin\theta\ .\label{gyY}
\end{align}
By comparing Eqs. (\ref{Y}) and (\ref{gyY}), taking into account our choice $a_Y=b_Y=1$, we get the following relations among the coupling constants $g_\alpha$, $g_\beta$, $g_Y$ and the mixing angle $\theta$:
\begin{align}
g_\alpha\cos\theta=g_\beta\sin\theta=\frac{g_Y}{\sqrt{2}}\ ,\label{gYg1g2theta1}
\end{align}
from which it follows that
\begin{align}
\tan\theta=\frac{g_\alpha}{g_\beta}\ \ \text{and}\ \ \frac{1}{g_\alpha^2}+\frac{1}{g^2_\beta}=\frac{2}{g^2_Y}\ .\label{gYg1g2theta2}
\end{align}
By changing to the basis defined by \eqref{secondBreaking}, the Lagrangian 
in \eqref{L2} can be rewritten as
\begin{align}
-\mathcal{L}\supset eJ^\mu_\gamma A_\mu+g_Z J^\mu_Z Z_{\mu} +g_{Z'}J^\mu_{Z'}Z'_\mu +g_WJ^\mu_{W^+} W^+_\mu + \text{h.c.} ,\label{L3}
\end{align}
where
\begin{align}
g_W=&\frac{g_L}{\sqrt{2}}\ , \notag\\
eJ^\mu_\gamma=& e\sum_{f}{\bar f_i}\gamma^\mu \hat{Q}(f_i)f_i\ , \notag\\
g_Z J^\mu_Z =&\frac{g_L}{2\cos\theta_W} \sum_{f}{\bar f_{i}}\gamma^\mu\left(\epsilon^L_{fi}P_L+\epsilon^R_{fi}P_R \right)f_{i}\ ,
\end{align}
with the chiral couplings to the $Z$ boson defined as
\begin{align}
\epsilon^{L,R}_{fi} = 2\left\lbrace \hat{T}_{3}(f_{iL,R})-\sin^2\theta_W\left[\hat{T}_{3}(f_{iL,R})+\frac{1}{2}\hat{Y}(f_{iL,R})\right]\right\rbrace.
\end{align}
To obtain these expressions, the identification $e=g_L\sin\theta_W=g_Y\cos\theta_W$ was made, which implies the well-known relation
\begin{align}
g_Y = g_L \tan\theta_W\ .\label{gYgLthetaW}
\end{align}
Taking into account the relations in Eqs. (\ref{gYg1g2theta1}) and (\ref{gYgLthetaW}), as well as the charges reported in Table \ref{tab:U1charges}, and the parameters $x$, $y$, $z$ and $w$ defined as
\begin{align}
    x\equiv&\ \frac{g_L\tan\theta_W}{2\sqrt{2}}\left(\cot\theta+\tan\theta\right)\alpha_{\nu 3}\ ,\notag\\
    y\equiv&\ \frac{g_L\tan\theta_W}{2\sqrt{2}}\left(\cot\theta+\tan\theta\right)\alpha_{\nu 1}\ ,\notag\\
    z\equiv&\ \frac{g_L\tan\theta_W}{2\sqrt{2}}\left[\cot\theta -3\left(\cot\theta+\tan\theta\right)\alpha_{q1}\right]\ ,\notag\\
    w\equiv&\ \frac{g_L\tan\theta_W}{2\sqrt{2}}\left(\cot\theta+\tan\theta\right)\alpha_{\sigma}\ ,\label{eq:xyzw}
\end{align}
the $Z'$ chiral couplings given by Eq. (\ref{gz'eLR}) can be expressed as indicated in Table \ref{Z'charges}\footnote{From here, we see that there are two families with identical charges and a different third one. However, universal models are still possible, for example, setting $z=x=y=1$ yields the well-known expressions for the $B-L$ charges.}. These charges are best suited for a phenomenological analysis of the new neutral vector boson, as it will be explained in the next section. Regarding the scalar fields $\Phi_1$ and $\Phi_2$, their $Z'$ couplings are given by $z-y$ and $z-x$, respectively.
\begin{table}[h!]
\begin{center}
%\begin{spacing}{1.5}
%\scalebox{0.8}{
%\tiny
\begin{tabular}{|c|c|c|c|c|c|c|c|c|}
\hline
$f$ & $\nu_{1,2}$  & $\nu_3$ & $e_{1,2}$  & $e_3$ &  $u_{1,2}$ & $u_3$ & $d_{1,2}$ & $d_3$\\ \hline 
$g_{Z'}\tilde\epsilon^L_f$&
      $-z$   & $-z$    &  $-z$  & $-z$  & $\frac{1}{3}z$   & $\frac{1}{3}z$  & $\frac{1}{3}z$  & $\frac{1}{3}z$   
\\
$ g_{Z'}\tilde\epsilon^R_f$ &
      $-y$  &  $-x$ &  $y-2z$ & $x-2z$  &  $-y+\frac{4}{3}z$  & $-x+\frac{4}{3}z$  &  $y-\frac{2}{3}z$ & $x-\frac{2}{3}z$
      \\   \hline
\end{tabular}
%}
%\end{spacing}
\caption{Chiral couplings between the fermion sector and the $Z^{\prime}$ gauge boson.}
\label{Z'charges}
\end{center}
\end{table}

\section{Low energy and collider constraints}

\begin{figure}%[h!]
%\vspace{-96pt}
\begin{center}
\centering 
\begin{tabular}{cc}
 \includegraphics[scale=0.3]{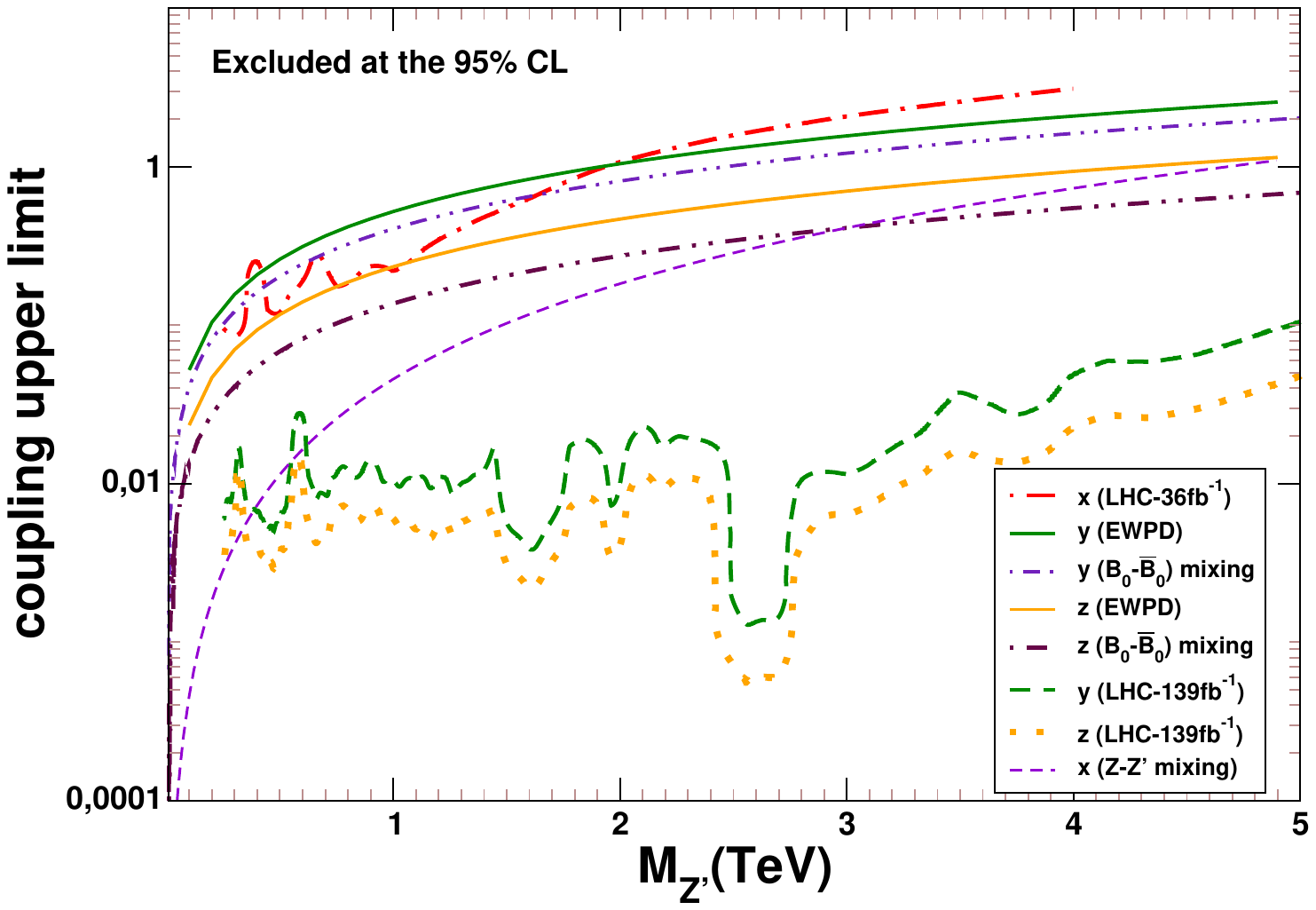}   &  \includegraphics[scale=0.3]{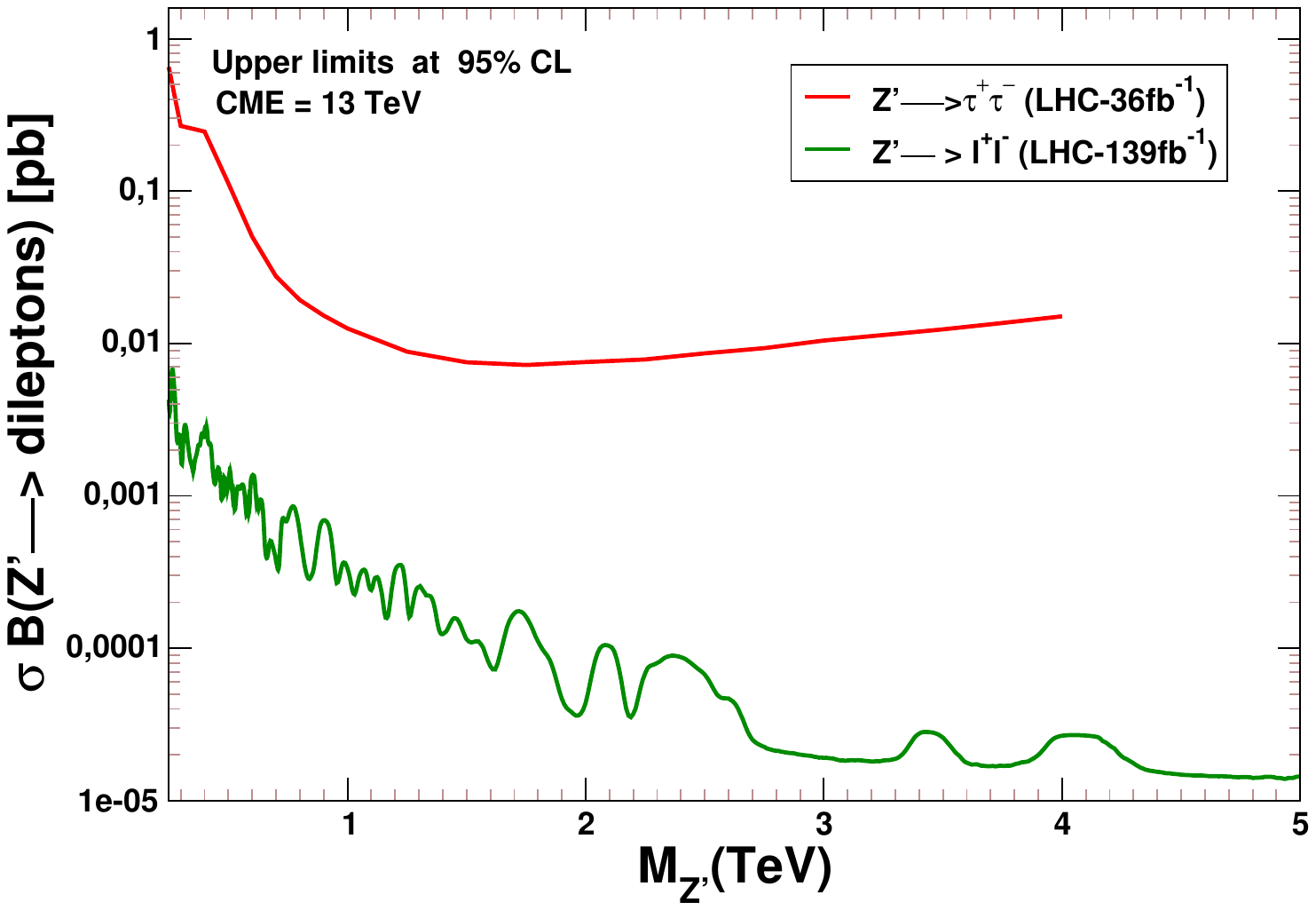}%\\   % & 
%&\includegraphics[scale=0.31]{minimal.pdf}% \\
%\includegraphics[scale=0.30]{minimal.pdf} \\
\end{tabular}
\end{center}
%\captionsetup{margin={1pt,0.05\textwidth}}
\caption{
Left:  Upper limit on the model parameters $x,y,z$. Right: 95\%  CL upper limits on the fiducial $Z'$ production  cross-section times the $Z'  \rightarrow \ell^+\ell^-$ branching~\cite{ATLAS:2019erb} (green continuous line)  and the corresponding upper limits on the $Z'$ decaying to $\tau\overline{\tau}$ pairs~\cite{ATLAS:2017eiz} (red continuous line).
}
\label{fig:1}	
\end{figure}

For the process $\bar{q}q\longrightarrow Z'\longrightarrow \ell^+\ell^-$, ATLAS reports upper limits on the fiducial cross-section times the  $Z'\rightarrow \ell^+\ell^-$  branching from searches of high-mass dilepton resonances (dielectron and dimuon) during Run 2 of the Large Hadron Collider~(LHC) at a center-of-mass energy of $\sqrt{s}=13$TeV and an integrated luminosity of 139fb$^{-1}$.  From these constraints, we obtain upper limits on 
the $y$  and $z$ couplings corresponding to the green dashed and orange dotted lines in the left-handed plot in Figure~\ref{fig:1}. These limits are obtained from the intersection of the theoretical cross-section
{(for further details see our previous publications~\cite{Erler:2011ud,Salazar:2015gxa,Rojas:2015tqa,Benavides:2018fzm})}
 with the 95\% CL upper limit on 
the cross-section reported by the ATLAS collaboration~\cite{ATLAS:2019erb}~(the  green continuous line in the right plot  Figure~\ref{fig:1}). 
For the upper limits on the $x$ parameter (red dot-dashed line in the left plot Figure~\ref{fig:1}), we use the ATLAS 95\%  upper limits on the production cross-section times branching fraction for a $Z'$ boson decaying to a $\tau\overline{\tau}$ pair ~(the green continuous line in the right plot Figure~\ref{fig:1}).  This data was collected by ATLAS in searches of $Z'$ bosons using a data sample corresponding to an integrated luminosity of 36.1fb$^{-1}$ from proton-proton collisions at a center of mass energy of  13 TeV~\cite{ATLAS:2017eiz}.

Constraints on a parameter are obtained by marginalizing the other parameters. In the case of the parameter $x$, which represents the coupling strength between $Z'$ and the fermions of the third family, the $Z'$ is produced from an annihilation $b\overline{b}->Z'$. 
Due to the strong collider constraints on the first two families, only $Z'$ couplings with the third family are possible at low energies.
In our model, this implies that $y,z\ll x$, as we can see from Table~\ref{Z'charges}. This implies that at low energies, the unique generator with unsuppressed coupling strengths is $T_{3R}(3)$. This symmetry is a well-known EW extension of the SM. The argument $``3"$ refers to the third family, and the subscript $3R$ refers to the generator $\sigma_{3}/2$, whose representation in the third family of the SM is $( b_R,t_R)^T$. If we allow right-handed neutrinos, as is, in fact,  the case in our model, a lepton representation $( \tau_{R},\nu_{\tau R})^T$ is also possible.

From reference~\cite{Erler:2009jh}, the $Z-Z'$ mixing angle $\Theta$ is restricted to be less than $10^{-3}$, which holds true for most models. Based on this result, we can assume $\Theta$ identically zero, which is a  typical assumption in collider constraints~\cite{Leike:1993ky}.

We also report Electroweak Precision data~(EWPD) constraints on the $y$ and $z$ parameters ( green and orange continuous lines in the left panel of Figure~\ref{fig:1}), obtained using the GAPP package~\cite{Erler:1999ug,Erler:2009jh},    which includes low-energy weak neutral current experiments ({this includes weak charges of the cesium atom and electron, as well as the constraints coming from cross-section ratios of neutrinos and antineutrino deep inelastic scattering. Measurements of the top and W masses are also in this set of observables}) and $Z$-pole observables.
%%\begin{widetext}
%\scalebox{0.8}{
%} % scalebox 
%\end{widetext}

As our model is non-universal, it has two possible sources of FCNC: the non-universal couplings of the $Z'$ and the couplings of the SM fermions to two scalar doublets.
Since the charges of the first two families are equal, we can ignore constraints from observables with flavor changes between quarks and leptons of the first two families, such as: $K^{0}$-$\bar{K}^{ 0}$-mixing, $\mu$-$e$ conversion, etc. In our case, one of the strongest constraints on the parameters comes from $B^{0}$-$\bar{B}^{0}$-mixing. Figure~\ref{fig:1} shows the upper limits on the  $y$ and $z$ parameters at a 95\% confidence level. 
{
An extended Higgs sector generates a mixture of $Z$ and $Z'$ that is proportional to the couplings of the Higgs to $Z'$ and to the expectation values of the neutral components of the scalar doublets. As shown in Appendix~\ref{sec:mixing}, the $Z-Z'$ constraints are relevant for the $x$ parameter because $z$ and $y$ are strongly constrained for colliders. As can be seen from the purple dashed line in Figure 1, this is  the most restrictive constraint on the $x$ parameter. }

In two Higgs doublet models, FCNC can be avoided if the mass matrix for SM fermions with the same electric charge and isospin is generated from a single Higgs doublet. As we show in the appendix~\ref{sec:sfcnc}, if the right-handed SM fermion is a singlet under the gauge group and if each right-handed SM singlet fermion couples to only one Higgs doublet (there is no problem if the scalar doublet has non-zero couplings to several right-handed fermions.), then there are no FCNC for the scalar sector; this is the case in our model.
%%%%%%%%%%%
\section{Analysis of Higgs-like resonant signals}
 Recently, several anomalies have been reported in
searches of high-mass scalar resonances in
proton-proton collisions at the LHC.  
The 2HDMs are the most straightforward extensions of the Standard Model that can explain these observations. Additionally, our model includes a scalar singlet $\sigma$ that gives mass to the $Z'$. The Higgs mechanism requires at least two CP-odd bosons to provide mass to the $Z$ and the $Z'$ and one charged scalar boson to give mass to the SM  $W$ boson, which leaves us with three CP even scalar bosons, one CP-odd scalar boson and one charged scalar. Our analysis aims to determine the typical masses for these bosons in the best-motivated parameter space and compare them with the experimental anomalies reported in the literature.
As explained in detail in the Appendix \ref{sec:potential}, of the three neutral scalar fields in the interaction space, $h_1$, $h_2$, and $\xi$, we can obtain using a unitary transformation, the neutral states in the mass space, $H_1$, $H_2$, $H_3$.
Great interest has generated an anomaly that can be explained by a light neutral scalar Higgs with a mass $M_{H1}\approx95$~GeV~\cite{CMS:2018cyk} and a charged Higgs around $M_{C ^{\pm}}\approx$~130~GeV~\cite{ATLAS:2023bzb}.
For the charged Higgs, in~\cite{Arhrib:2024sfg} a detailed analysis of the phenomenological implications of a new resonance with a three sigma significance was studied.
On a mass basis, we will denote as $M_A$ the only CP-odd field that is not absorbed as a Goldstone boson.
An excess of events was also found in channels involving the productions of SM gauge bosons, $\gamma\gamma$ and $Z\gamma $ (for further analysis, look in~\cite{Crivellin:2021ubm} and references therein). This analysis provides a good indication of new scalar resonances decaying into two photons with invariant masses of 95 GeV~\cite{Banik:2023ecr} and 152 GeV~\cite{Crivellin:2021ubm}. 
Other excesses over the expected value in the SM for dibosons are reported at 680 GeV~\cite{CMS:2017dib}, which are compatible with the excess in $\gamma\gamma$ and $b\bar b$ reported by the CMS collaboration~\cite{CMS:2022tgk}. A more complete review of these anomalies and additional references can be found at~\cite{Crivellin:2023zui}. In this reference, they also mention an excess reported by the ATLAS collaboration that can be interpreted as a pseudoscalar with a mass of 650~GeV produced in association with a scalar with a mass of 450~GeV.
  
Recently, a deviation from the background-only expectation occurred for high scalar resonances with masses $(575, 200)$ GeV and a local (global) significance of 3.5 (2.0) standard deviations, as reported by the ATLAS collaboration~\cite{ATLAS:2024auw}.
It is important to stress that this analysis shows good agreement with the background-only hypothesis for the masses $(650, 90)$ GeV, where CMS reported an excess with a local (global) significance of 3.8 (2.8) standard deviations~\cite{CMS:2022tgk}. 
  
To account for these experimental anomalies from the scalar potential of our model (see equation~\ref{pot}), we consider two possible assignments for the charges of the scalar singlet $\sigma$. One of them leads to a cubic coupling among the scalar fields, while the other to quartic term~(the remaining terms in the scalar potential~\ref{pot} are always present regardless of the $\sigma$ charge). If the $Z'$ coupling of $\sigma$ is $x-y$, that is, $\alpha_\sigma = \alpha_{\nu 3}-\alpha_{\nu 1}$, then the following term is allowed:
\begin{align*}
\mu\left[(\Phi_1^\dag\Phi_2)\sigma+\text{h.c.}\right]   \ .
\end{align*}
In this case, the coupling
constant $\mu$ has dimensions
of mass, and in order to
have a consistent mass spectrum, its values must be in the range
%\begin{equation}
$-77.3~\text{GeV}\leq \mu < 0 $. 
%\label{F1}  
%\end{equation}
Similarly, if the coupling of $\sigma$ to the $Z'$ boson is $\frac{1}{2}(x-y)$, or equivalently $\alpha_\sigma = (\alpha_{\nu 3}-\alpha_{\nu 1})/2$, it is possible to form the term
\begin{align}
\lambda\left[(\Phi_1^\dag\Phi_2)\sigma^2+\text{h.c.})\right],
\end{align}
where the constant $\lambda$ is dimensionless, and restricted to the range $(-0.44,0)$.
According to our scalar 
sector (whose scalar 
potential we show
in Appendix A), 
to reproduce part of the spectrum of anomalies in the scalar sector~(determined by the VEVs and coupling constants) we must 
identify the middle-mass neutral scalar boson 
as the SM Higgs boson to 
which we assign its 
well-known mass of 
$M_{H_2}=125$~GeV. 
In our model, we ensure that only the massive charged scalar field coincides with the anomaly $M_{C^\pm}=130$~GeV. This happens while the SM vector boson $W$ absorbs the other massless-charged field through the Higgs mechanism.
Similarly, the Higgs mechanism requires a pseudoscalar field from one of the scalar doublets to give mass to the $Z$ boson, and the pseudoscalar field of the scalar singlet to give mass to the $Z'$.
 The mass of the remaining scalar fields ($M_{H_1}$,  $M_{H_3}$  and  $M_{A}$) are free parameters.
For the other dimensionless parameters of the potential~\eqref{pot}, their values are assumed to be in the range $[-1.5,1.5]$. 
Regarding the VEVs of the $\Phi_1$ and $\Phi_2$, they are chosen such that $v_1^2+v_2^2=(246.24~\text{GeV})^2$ with $v_2\gg v_1$. This hierarchy between VEVs is necessary to align the Higgs doublet $\Phi_2$ with that of the SM.
We take the VEV of the scalar singlet $\langle \sigma\rangle= v_\sigma/\sqrt{2}$ as a free parameter varying between 250 GeV and 2000 GeV. Finally, in order to satisfy the collider constraints, we require $z,y \ll x$, and take $x\gtrsim 1$ for $Z'$ masses above $2$ TeV,  as explained in section~\ref{sec:gauge}.  

\begin{figure}%[h!]
%\vspace{-96pt}
\begin{center}
 \includegraphics[scale=0.4]{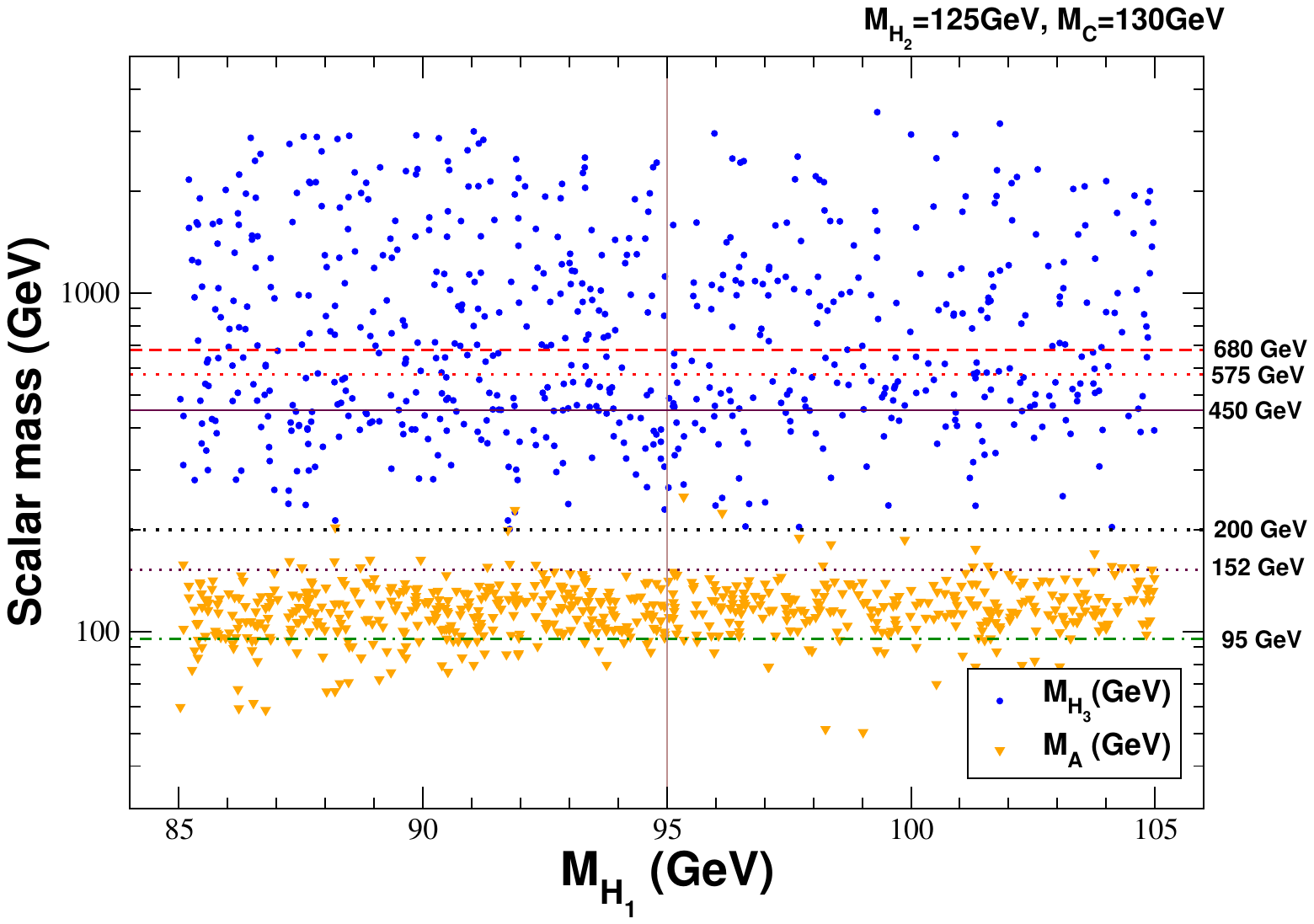}
\end{center}
%\captionsetup{margin={1pt,0.05\textwidth}}
\caption{
Distribution of the scalar mass $M_{H_3}$ (the blue round points) and the pseudoscalar $M_{A}$ (orange triangle points) for a scalar potential including the cubic term $\mu\Phi_1^\dagger\Phi_2\sigma+\text{h.c.}$.
In this term, $\mu$ has mass dimensions and takes values in the range $(-77.3,0)$~GeV.
We vary the other 
dimensionless 
parameters of 
the scalar potential 
in the range $[-1.5, 
1.5]$, and take 
the VEV of the 
scalar singlet $v_\sigma$ as a free 
parameter ranging 
from 250 to 
2000~GeV.
The VEV's  $v_1$ and $v_2$ vary subject to the conditions $\sqrt{v_1^2+v_2^2}=v=246.24~\text{GeV}$ and $v_2\gg v_1$.
The masses of the scalars and pseudo-scalars: $M_{H_1}$, $M_{H_3}$ and $M_{A}$ are determined by the tadpole equations~\eqref{ligaduras1}. See the text for further details (see appendix~\ref{sec:potential}). }
\label{fig:cubico}	
\end{figure}

\begin{figure}%[h!]
%\vspace{-96pt}
\begin{center}
 \includegraphics[scale=0.4]{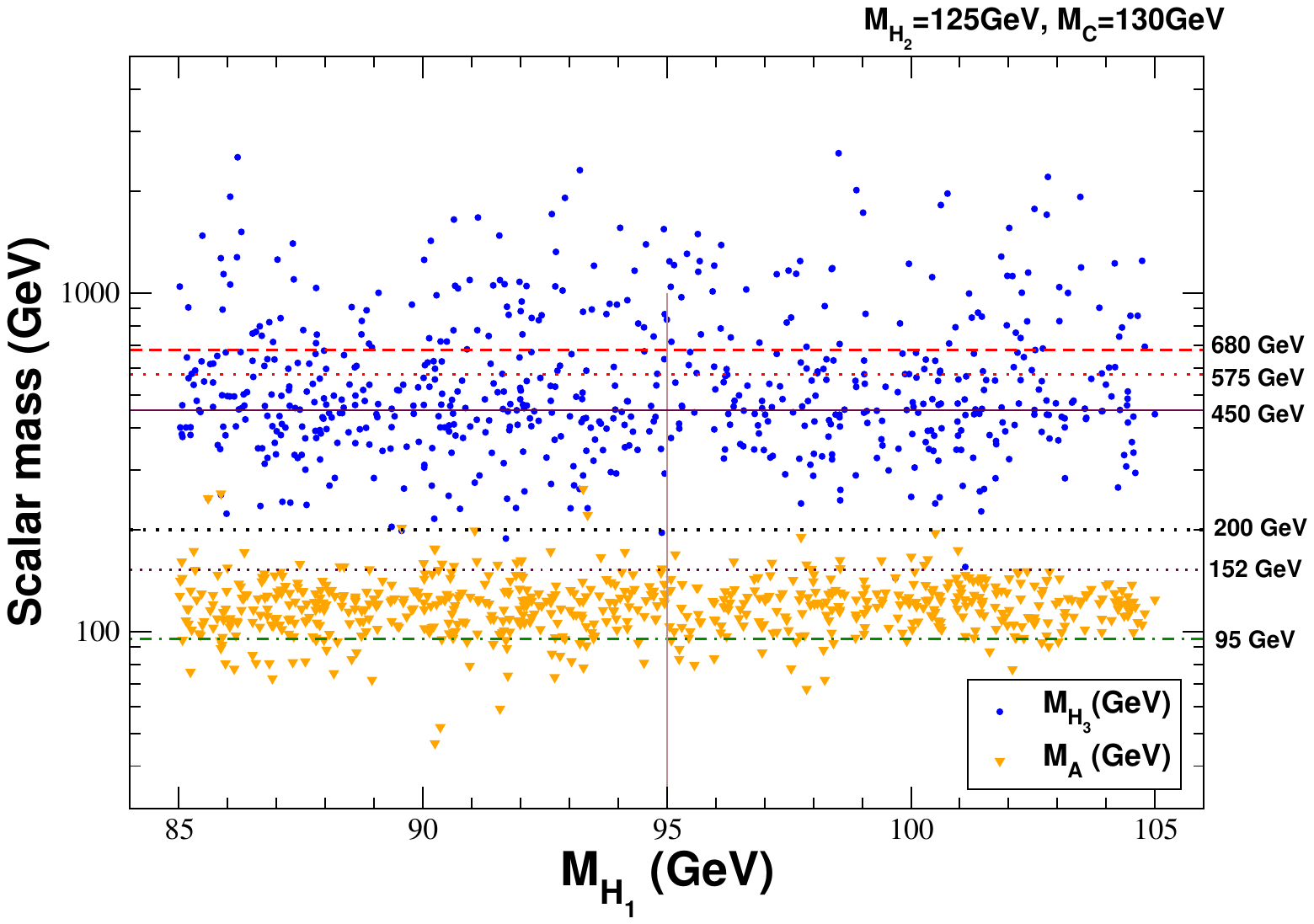}
\end{center}
%\captionsetup{margin={1pt,0.05\textwidth}}
\caption{
Distribution of the scalar mass $M_{H_3}$ (the blue round points) and the pseudo-scalar $M_{A}$ (orange triangle points) for a scalar potential including the quartic term $\lambda\Phi_1^\dagger\Phi_2\sigma^2+\text{h.c.}$.
In this term, $\lambda$ is dimensionless and takes values in the interval $(-0.44,0)$.
We vary the other 
dimensionless 
parameters of 
the scalar potential 
in the range $[-1.5, 
1.5]$, and take 
the VEV of the 
scalar singlet $v_\sigma$ as a free 
parameter ranging 
from 250 to 
2000~GeV.
The VEV's  $v_1$ and $v_2$ vary subject to the conditions $\sqrt{v_1^2+v_2^2}=v=246.24~\text{GeV}$ and $v_2\gg v_1$.
The masses of the scalars and pseudoscalars: $M_{H_1}$, $M_{H_3}$ and $M_{A}$ are determined by the tadpole equations~\eqref{ligaduras1}. See the text for further details. }
\label{fig:cuartico}	
\end{figure}

To illustrate the density of solutions, Figures \ref{fig:cubico} and \ref{fig:cuartico} display a total of 640 solutions spread across the $M_{H_1}$ vs $M_{H_3}$ and $M_{H_1}$ vs $M_A$ axes. It is important to emphasize our identification of the lightest CP-even Higgs scalar $H_1$ with the anomaly at 95 GeV. Therefore, we have considered exploring the mass interval 95 $\pm$ 10 GeV.
In these figures, we can see that many of the experimental anomalies coincide with the regions with the highest density of solutions. This coincidence is important since we have made the free parameters of the theory vary in intervals that we consider natural.
{
It is very important that in our analysis we imposed the hierarchy $m_{H_1}<m_{H_2}<m_{H_3}$. The results are identical if we choose $m_{H_2}<m_{H_1}$, and identify the Higgs with $M_{H_1}$. From the density plot, we see that the highest density of solutions for the pseudo-scalar mass is between 100GeV and 200GeV. This result holds for both cubic and quartic potentials.
In this range, we have three experimental anomalies (the one at 95GeV, the one at 152GeV, and the one at 200GeV). If any of these anomalies accumulate statistics, this strongly suggests that a pseudoscalar particle could explain the resonance.

For the quartic potential (see Figure~\ref{fig:cuartico}), the highest density of solutions is found below 1000 GeV, while for the cubic potential (see Figure~\ref{fig:cubico}), there is a high density of solutions up to 3000 GeV.
Below 1000 GeV we have three experimental anomalies with masses: 450 GeV, 575 GeV, and 680. From the density plots we see that for a quartic potential, it is more probable to have solutions in this range when compared to the cubic potential.
}

\section{Conclusions}

In this work, we assume that the SM is a low-energy effective theory of a more fundamental theory characterized by a gauge symmetry of the form $SU(3)_\text{C}\otimes SU(2)_L\otimes  U(1)_\alpha \otimes U(1)_\beta$, and whose particle content is that of the SM extended with three right-handed neutrinos, a second Higgs doublet and a scalar singlet. Additionally,  we impose that both $U(1)$ charges are non-universal and contribute non-trivially to the  SM hypercharge, i.e., they are not inert charges.  Under these assumptions, we showed that the most general expression for the $Z'$ chiral couplings  is as those shown in  Table~\eqref{Z'charges}. In this model, it is possible to generate all the mass matrix elements of  with only two Higgs doublets. From this, it is possible to adjust the model to reproduce the CKM and PMNS mixing matrices. This feature is highly non-trivial for non-universal scenarios and represents a great advantage of this model. It is important to mention that to maintain the non-universality condition, it was preferable to avoid Majorana mass terms 
{ (if we want to reproduce the electric charges of the SM particles (which are universal) from non-universal $U(1)$ charges, in most of the cases studied, we must avoid introducing neutrino Majorana masses).}

From the assumptions of our work, as well as the collider, electroweak and flavor constraints, we also conclude that for a model with two non-inert Abelian symmetries at low energies ($M_{Z'}< 5$~TeV), only the residual symmetry $T_{3R}(3)$, in addition to the SM gauge symmetry, has an unsuppressed coupling strength. The argument $``3"$  says that only couplings to the third family are possible. Models with couplings to the first and second families are strongly constrained, so that only $Z'$ couplings below 0.1 are possible, i.e.,  $g_{Z'} \tilde\epsilon_{L,R}<0.1$. For a $Z'$ coupling to the third family, it is possible to have $Z'$ charges such that  $g_{Z'}\tilde\epsilon_{L,R}\sim 1$ for $Z'$ masses above 2 TeV.

Our work analyzes some Higgs-like anomalies recently reported by the ATLAS and CMS collaborations~\cite{Crivellin:2021ubm}.
To this end, we show the distribution of 400 solutions in the $M_{H_1}$, $M_{H_3}$ and $M_{H_1}$, $M_{A}$ planes. These results are shown in Figures~\ref{fig:cubico} and \ref{fig:cuartico}. This analysis concludes that explaining some of the observed anomalies within the model is possible.

We show that  the scalar sector FCNC cancel if each right-handed fermion couples only to a single Higgs doublet (although the scalar doublet can have non-zero couplings with several right-handed fermions). This will be the case as long as the right-handed fermions are singlets of the gauge group.

\appendix

\section*{Appendices}

\section{Scalar potential~\label{sec:potential}}
Our model contains two scalar doublets, $\Phi_1$  and $\Phi_2$, and a scalar singlet $\sigma$. In general, these fields can be expressed as  
\begin{equation}
\Phi_1=\begin{pmatrix}
\phi_1^+\\
\frac{v_1+h_1+i\,\eta_1}{\sqrt{2}}
\end{pmatrix},\hspace{1cm}
\Phi_2=\begin{pmatrix}
\phi_2^+\\
\frac{v_2+h_2+i\,\eta_2}{\sqrt{2}}
\end{pmatrix},\hspace{1cm}
\sigma=
\frac{v_\sigma+\xi+i\,\zeta}{\sqrt{2}} \ ,
\label{escalares}
\end{equation}
where $\langle\Phi_1\rangle=(0,v_1/\sqrt{2})^T$, $\langle\Phi_2\rangle=(0,v_2/\sqrt{2})^T$  and $v_\sigma=\sqrt{2}\langle\sigma\rangle$. 
For the doublet $\Phi_2$ (which is close to $H_1$ in the Georgi basis) to be aligned with the Higgs of the SM  we impose the hierarchy
\begin{align}
v_{\sigma}> v_2\gg v_1\ .
\label{jerarquia} 
\end{align}
Since the Higgs doublet is a linear combination of the two scalar doublets, then
\begin{equation}
\sqrt{v_1^2+v_2^2}=v=246.24~\text{GeV} \ ,
\end{equation}
The most general scalar potential consistent with the gauge symmetry $SU(2)_L\otimes U(1)_\alpha \otimes U(1)_\beta$ is~\cite{Banik:2023ecr}:
\begin{equation}
\begin{split}
V(\Phi_1,\Phi_2,\sigma)&= \mu_1^2\,|\Phi_1|^2+\mu_2^2\,|\Phi_2|^2+\mu_\sigma^2\,|\sigma|^2+
\lambda_1\,|\Phi_1|^4+\lambda_2\,|\Phi_2|^4+\lambda_\sigma\, |\sigma|^4\\
&+\lambda_{3}\,|\Phi_1|^2|\Phi_2|^2 +\lambda_{4}\,|\Phi_1^\dag\Phi_2|^2+ \lambda_{1\sigma}\,|\Phi_1|^2|\sigma|^2+\lambda_{2\sigma}\,|\Phi_2|^2|\sigma|^2\\
&
+\text{linear term in }\sigma\ 
(\text{or quadratic term in } \sigma )\ ,
\end{split}
\label{pot}
\end{equation}
where a linear interaction term in $\sigma$ (which we will denote as the cubic term) of the form
\begin{align*}
\mu\left[(\Phi_1^\dag\Phi_2)\sigma+\text{h.c.}\right] 
\end{align*}
is possible if $\alpha_\sigma$ in Table \ref{tab:U1charges} is taken to be $\alpha_{\nu 3}-\alpha_{\nu 1}$. Here $\mu$ is a coupling with mass dimensions. On the other hand, if $\alpha_\sigma$ is equal to $\frac{1}{2}(\alpha_{\nu 3}-\alpha_{\nu 1})$, then the quadratic term in $\sigma$
(which we will denote as the quartic term),
\begin{align*}
\lambda\left[(\Phi_1^\dag\Phi_2)\sigma^2+\text{h.c.})\right]\ ,    \end{align*}
is the one that is present. In this case, the coupling $\lambda$ is dimensionless. 
By minimizing the potential in Eq.~\eqref{pot}, we then  obtain  that   
%Al minimizar el potencial~\eqref{pot},
%
\begin{equation}
\begin{split}
\mu_1^2&=-\frac{\sqrt{2}}{2}\frac{\mu v_2 v_\sigma}{v_1} -\lambda_1 v_1^2 - \frac{\lambda_3+\lambda_4}{2}v_2^2  - \frac{\lambda_{1\sigma}}{2} v_\sigma^2\ ,\\
\mu_2^2&=-\frac{\sqrt{2}}{2}\frac{\mu v_1 v_\sigma}{v_2} -\lambda_2 v_2^2- \frac{\lambda_3+\lambda_4}{2}v_1^2  - \frac{\lambda_{2\sigma}}{2} v_\sigma^2\ ,\\
\mu_\sigma^2&=-\frac{\sqrt{2}}{2} \frac{\mu v_1 v_2}{v_\sigma }-\lambda_\sigma  v_\sigma^2 -\frac{\lambda_{1\sigma}}{2} v_1^2 - \frac{\lambda_{2\sigma}}{2} v_2^2 \ .
\end{split}
\label{ligaduras1}
\end{equation}
in the cubic case, while in the quartic one, the corresponding expressions can be obtained from the previous ones by making the substitution $\sqrt{2}\mu\rightarrow \lambda v_\sigma$\ .
%
%
%%%%%%%%%%%
\subsection{Mass spectrum of the neutral scalar sector}
From the potential~\eqref{pot} and the previous minimization conditions, we can build the mass matrices from the fields defined in Eq. \eqref{escalares}. For the  CP-even scalar field basis $(h_1, h_2, \xi)$, the mass matrix is given in the cubic case~\cite{Ponce:2002sg} by:  
\begin{equation}
\begin{pmatrix}
2 \lambda_1 v_1^2-\frac{\mu v_2 v_\sigma}{\sqrt{2} v_1}&\frac{\mu v_\sigma}{\sqrt{2}}+v_1 v_2 (\lambda_3+\lambda_4)&\frac{\mu v_2}{\sqrt{2}}+\lambda_{1\sigma} v_1 v_\sigma\\
\frac{\mu v_\sigma}{\sqrt{2}}+v_1 v_2 (\lambda_3+\lambda_4)&
2 \lambda_2 v_2^2-\frac{\mu v_1 v_\sigma}{\sqrt{2} v_2}&
\frac{\mu v_1}{\sqrt{2}}+\lambda_{2\sigma} v_2 v_\sigma\\
\frac{\mu v_2}{\sqrt{2}}+\lambda_{1\sigma} v_1 v_\sigma&\frac{\mu v_1}{\sqrt{2}}+\lambda_{2\sigma} v_2 v_\sigma&2 \lambda_\sigma  v_\sigma^2-\frac{\mu v_1 v_2}{\sqrt{2} v_\sigma}
\end{pmatrix},
\label{mmescalar}
\end{equation}
while in the quartic case, it corresponds to:
\begin{equation}
\begin{pmatrix}
 2 \lambda_1 v_1^2-\frac{\lambda v_2 v_\sigma^2}{2 v_1} & \frac{\lambda v_\sigma^2}{2}+v_1 v_2 (\lambda_3+\lambda_4) & v_\sigma (\lambda v_2+\lambda_{1\sigma} v_1) \\
 \frac{\lambda v_\sigma^2}{2}+v_1 v_2 (\lambda_3+\lambda_4) & 2 \lambda_2 v_2^2-\frac{\lambda v_1 v_\sigma^2}{2 v_2} & v_\sigma (\lambda v_1+\lambda_{2\sigma} v_2) \\
 v_\sigma (\lambda v_2+\lambda_{1\sigma} v_1) & v_\sigma (\lambda v_1+\lambda_{2\sigma} v_2) & 2 \lambda_{\sigma}  v_\sigma^2
\end{pmatrix}.
\label{mmescalar2}
\end{equation}
These are square mass
matrices of rank three
 with mass eigenvalues $M_{H_1}$, $M_{H_2}$ and $M_{H_3}$, corresponding to the mass eigenstates $H_1$, $H_2$ and $H_3$, respectively. We will identify the states according to the mass hierarchy:
\[
M_{H_1}< M_{H_2}<M_{H_3}\ .
\]
%
%%%%%%%%%%%
The intermediate-mass scalar state, $H_2$, can be identified as the SM Higgs, while the light mass scalar state $H_1$ and the heavy mass scalar state $H_3$  are new scalar fields that,     in principle, can be observed in the LHC experiments. The hierarchy~\eqref{jerarquia} causes the  scalar $H_2$ to align with $h_2$.
%%%%%%%%%%%
\subsubsection{Mass spectrum of the neutral pseudoscalar sector}
In the $(\eta_1, \eta_2, \zeta)$ basis, the pseudoscalar squared mass matrix takes the following form for the cubic case:
\begin{equation}
\frac{\mu}{\sqrt{2}}\begin{pmatrix}
-\dfrac{v_2 v_\sigma}{v_1}&v_\sigma & v_2\\
 v_\sigma&-\dfrac{v_1 v_\sigma}{v_2}&- v_1\\
v_2 &-v_1&-\dfrac{v_1 v_2}{v_\sigma}
\end{pmatrix} .\label{mmpseudoscalar1}
\end{equation}
%
%Para el caso cuártico tenemos:
The corresponding mass matrix for the quartic case is 
\begin{equation}
\frac{\lambda v_\sigma}{2} \begin{pmatrix}
 -\dfrac{v_2 v_\sigma}{v_1} & v_\sigma &  2v_2  \\
 v_\sigma & -\dfrac{ v_1 v_\sigma}{ v_2} & - 2v_1  \\
 2v_2  & -2v_1  & -4 \dfrac{v_1 v_2}{v_\sigma} \\
\end{pmatrix}.\label{mmpseudoscalar2}
\end{equation}
In both cases, these mass matrices have rank  1. The two zero eigenvalues correspond to the two Goldstone bosons that give mass to the $Z$ and $Z^\prime$ bosons after the spontaneous symmetry breaking. The non-zero eigenvalue corresponds to a measurable pseudoscalar with mass equal to:
\begin{align}
    M^2_A=\begin{cases}
    -\dfrac{\mu \left(v_1^2 v_2^2+v^2v_\sigma^2\right)}{\sqrt{2} v_1 v_2 v_\sigma}\,\,\,\,\,\,\, (\text{cubic case})\ ,\\\\
    -\dfrac{\lambda \left(4 v_1^2 v_2^2+v^2 v_\sigma^2\right)}{2 v_1 v_2}\,\,\,\, (\text{quartic case})\ ,
    \end{cases}
\end{align}
whose mixing comes mainly from $\eta_1$.
%%%%%%%%%%%
%%%%%%%%%%%
\subsubsection{Mass spectrum of the charged scalar sector}
In the $(\phi_1^\pm, \phi_2^\pm)$ basis, the squared mass matrix for charged scalar particles is
\begin{equation}
\frac{1}{2}\begin{pmatrix}
-\sqrt{2}\dfrac{\mu v_2 v_\sigma}{v_1} -\lambda_4 v^2_2  &  \sqrt{2} \mu v_\sigma+\lambda_4 v_1 v_2\\
\sqrt{2} \mu v_\sigma+\lambda_4 v_1 v_2 &-\sqrt{2}\dfrac{\mu v_1 v_\sigma}{v_2} -\lambda_4 v^2_1
\end{pmatrix}, \label{mmchargedscalar1}
\end{equation}
for the cubic case, and
\begin{equation}
\frac{1}{2}\begin{pmatrix}
 -\dfrac{\lambda v_2 v_\sigma^2}{v_1}-\lambda_4 v_2 & \lambda v_\sigma^2+\lambda_4 v_1 v_2\ \\
 \lambda v_\sigma^2+\lambda_4 v_1 v_2 & -\dfrac{\lambda v_1 v_\sigma^2}{v_2}-\lambda_4 v_1 
\end{pmatrix},\label{mmchargedscalar2}
\end{equation}
for the quartic case. As before, these mass matrices have rank 1, with the only zero eigenvalue corresponding to the
Goldstone boson giving mass to the charged $W$ boson. The remaining charged scalar acquires a mass equal to 
\begin{align}
M^2_{C^\pm}=\begin{cases}
-\dfrac{v^2}{2}\left(\sqrt{2} \dfrac{\mu v_\sigma}{v_1 v_2} +\lambda_4 \right)\ ,\,\,\,\,\,(\text{cubic case})\ ;\\\\ 
-\dfrac{v^2}{2}\left(\dfrac{\lambda v_\sigma^2}{v_1 v_2} +\lambda_4 \right)\ ,\,\,\,\,\,\,\,\,\,\,\,\, (\text{quartic case})\ .
\end{cases}
\end{align}

\section{The gauge boson masses \label{sec:gauge}}

Let us now determine the mass of the neutral gauge bosons. These are obtained from the scalar-gauge couplings introduced by the covariant derivatives of the scalar fields in the Lagrangian terms 
\begin{align}
\mathcal{L}\supset |D_\mu \Phi_1|^2 + |D_\mu \Phi_2|^2 + |D_\mu \sigma|^2\ ,
\end{align}
where
\begin{align}
    D^\mu =\partial^\mu +\frac{i}{2}g_L A^\mu_j\hat{T}_j+\frac{i}{2}g_\alpha B^\mu_\alpha \hat{\alpha} +\frac{i}{2}g_\beta B^\mu_\beta \hat{\beta}\ ,
\end{align}
with $\hat{T}_j$, $A^\mu_{j}$ ($j = 1,2,3$) and $g_L$ denoting, respectively, the generators, the gauge fields and the coupling constant associated with the weak isospin gauge group $SU(2)_L$\footnote{The $SU(2)_L$ generators are defined in terms of the Pauli matrices according to
%\begin{align}
$    T_i=\frac{1}{2}\sigma_i\ . 
$ 
%\end{align}
}, while $\hat{\chi}$, $B^\mu_\chi$ and $g_\chi$, with $\chi=\alpha,\beta$, are the corresponding quantities related with the two Abelian $U(1)$ factors. For the Higgs doublets $\Phi_a$ ($a=1,2$) and the singlet $\sigma$, we have
\begin{align}
    D^\mu\Phi_a &=\left[\partial^\mu +\frac{i}{2} g_L \begin{pmatrix}
           A^\mu_3 &\sqrt{2}W^\mu\\
           \sqrt{2}W^{\mu\dagger} &-A^\mu_3
           \end{pmatrix}+\frac{i}{2}g_\alpha \alpha_{a} B^\mu_\alpha +\frac{i}{2}g_\beta \beta_{a} B^\mu_\beta  \right]\Phi_a\ , \notag\\
    D^\mu\sigma &=\left(\partial^\mu +\frac{i}{2}g_\alpha \alpha_\sigma B^\mu_\alpha  + \frac{i}{2}g_\beta \beta_\sigma B^\mu_\beta \right)\sigma\ ,
\end{align}
Here $\alpha_a$ ($\beta_a$) and $\alpha_\sigma$ ($\beta_\sigma$) denote, respectively, the $U(1)_{\alpha(\beta)}$ charges for $\Phi_a$ and $\sigma$ given in Tab. \ref{tab:U1charges}. Additionally, the $W$ field has been defined as
\begin{align}
 W^\mu = \dfrac{A^\mu_1 - i A^\mu_2}{\sqrt{2}}\ .
\end{align}
Taking into account the definition of $\Phi_a$ and $\sigma$ given in Eq. (\ref{escalares}), as well as the basis changes defined in Eqs. (\ref{firstBreaking}) and (\ref{secondBreaking}), which imply
\begin{align}
    B^\mu_\chi &=a_\chi A^\mu + b_\chi Z^\mu +c_\chi Z'^\mu\ ,\ \ \ (\chi=\alpha,\beta)\ ; \notag\\
    A^\mu_3 &= \sin\theta_W A^\mu +\cos\theta_W Z^\mu\ ,
\end{align}
with
\begin{align}
    a_\alpha &=\cos\theta\cos\theta_W\ ,\ \ \  b_\alpha =-\cos\theta\sin\theta_W\ ,\ \ \ c_\alpha = -\sin\theta\ , \notag\\
    a_\beta &=\sin\theta\cos\theta_W\ ,\ \ \  b_\beta =-\sin\theta\sin\theta_W\ ,\ \ \ \ c_\beta = \cos\theta\ ,
\end{align}
it can be shown that the mass terms for the $Z^\mu$, $Z'^\mu$ and $W^\mu$ gauge bosons are
\begin{align}
    \mathcal{L}\supset &\frac{1}{2}\left(g^2_Z v^2\right)Z_\mu Z^\mu + 
    \frac{1}{2}\left[g^2_{Z'}\left(\gamma'^2_1v^2_1+\gamma'^2_2v^2_2+\gamma'^2_\sigma v^2_\sigma\right)\right]Z'_\mu Z'^\mu \notag\\
    &-\frac{1}{2}\left[2g_Z g_{Z'}\left(\gamma'_1v^2_1+\gamma'_2v^2_2\right)\right]Z_\mu Z'^\mu + g^2_W v^2 W^\dagger_{\mu}W^{\mu},
\end{align}
the coupling constants $g_{Z}$ and $g_{W}$ are defined as in the SM, \ie
\begin{align}
    g_Z=\frac{g_L}{2\cos\theta_W}\ ,\ \ \ g_{W}=\frac{g_L}{2}\ ,
\end{align}
while $g_{Z'}$ is defined through the following relations:
\begin{align}
    g_{Z'}\gamma'_a &=-\frac{1}{2}\left(g_\alpha \alpha_{a}\sin\theta - g_\beta \beta_{a}\cos\theta\right)\ ,\ \ \ (a=1,2.);\notag\\
    g_{Z'}\gamma'_\sigma &=-\frac{1}{2}\left(g_\alpha \alpha_{\sigma}\sin\theta - g_\beta \beta_{\sigma}\cos\theta\right).
\end{align}
In terms of the $x$, $y$, $z$ and $w$ parameters defined in Eq. (\ref{eq:xyzw}), these couplings can be expressed as
\begin{align}
    g_{Z'}\gamma'_1 =z-y\ ,\ \ \ g_{Z'}\gamma'_2 =z-x\ , \ \ \  g_{Z'}\gamma'_\sigma =-w\ .
\end{align}
Writing the $Z-Z'$ mixing matrix as
\begin{align}
    M^2_{Z-Z'}=\begin{bmatrix}
        g^2_Z v^2 &-g_Z g_{Z'}\gamma'_a v^2_a \\
        -g_Z g_{Z'}\gamma'_a v^2_a &g^2_{Z'}\left(\gamma'^2_av^2_a+\gamma'^2_\sigma v^2_\sigma\right)
    \end{bmatrix}\equiv
    \begin{pmatrix}
        \mathcal{A} &-\mathcal{C}\\
        -\mathcal{C} &\mathcal{B}\\
    \end{pmatrix},
\end{align}
with a sum over the $a$ index implied, then the square masses of the physical neutral gauge bosons $Z_1$ and $Z_2$ are given by
\begin{align}
    m^2_{Z_{1,2}}=\frac{1}{2}\left[ \mathcal{A} + \mathcal{B} \mp\sqrt{\left(\mathcal{A} -\mathcal{B} \right)^2+4 \mathcal{C}^2}\right].
\end{align}
If $\mathcal{O}$ is the diagonalizing orthogonal matrix defining the mass basis, \ie 
\begin{align}
\begin{pmatrix}
Z^\mu\\
Z'^\mu
\end{pmatrix}=
\mathcal{O}
\begin{pmatrix}
Z^\mu_1\\
Z^{\mu}_2
\end{pmatrix}
=
\begin{pmatrix}
\cos \Theta &-\sin \Theta\\
\sin \Theta &\cos \Theta
\end{pmatrix}
\begin{pmatrix}
Z^\mu_1\\
Z^{\mu}_2
\end{pmatrix},\label{Z1Z2bosons}    
\end{align}
then the mixing angle $\Theta$ can be determined from
\begin{align}
    \tan 2\Theta =\frac{2\mathcal{C}}{\mathcal{B}-\mathcal{A}}\ .
\end{align}
From this expression, it is possible to obtain 
\begin{align}
     \Theta =\frac{1}{2}
     \arctan\left\lbrace\frac{g_L}{2\cos\theta_W}
     \frac{
     \left[(z-y)v_1^2+(z-x)v_2^2\right]}
     {
     \left[(z-y)^2v_1^2+(z-x)^2v_2^2+w^2v_\sigma^2\right]
     -\frac{g_L^2v^2}{4\cos^2\theta_W}
     }\right\rbrace\ .
\end{align}
To satisfy the current constraint on the mixing angle~\cite{Erler:2009jh} it is necessary to keep this angle below $10^{-3}$, which is possible in two scenarios: 1)  a light $Z'$ mass, i.e., $M_{Z'}\ll M_{Z}$ or a heavy $Z'$  mass, i.e.,   $M_{Z'}\gg M_{Z}$, which requires $x\gtrsim 1$,  $z\sim y \ll 1$ and $v<v_\sigma $. { As usual in calculating collider constraints~\cite{Leike:1993ky} we assume $~\theta_{Z-Z'}=0$}. 
In analyzing the scalar anomalies $w=x-y$ in the cubic case, or $w=\frac{x-y}{2}$ for a potential with quartic coupling term (as explained in appendix~\ref{sec:potential}). 
For that analysis, assuming a heavy $Z'$ mass is more convenient.

\section{Analysis of scalar FCNCs \label{sec:sfcnc}}

The Yukawa interactions are described by the general Lagrangian
\begin{align}
-\mathcal{L}_Y={\bar q'_{iL}}y^{a d}_{ij}\Phi_a d'_{jR}+{\bar q'_{iL}}y^{a u}_{ij}\tilde{\Phi}_a u'_{jR}+{\bar l'_{iL}}y^{a e}_{ij}\Phi_a e'_{jR}+{\bar l'_{iL}}y^{a \nu}_{ij}\tilde{\Phi}_a \nu'_{jR}+ \text{h.c.}\ ,
\end{align}
with $i,j=1,2,3$, and $a=1,2$. Here
\begin{align}
\Phi_{a}=\begin{pmatrix}
\phi_{a}^+\\
\dfrac{v_a + h_a + i \eta_a}{\sqrt{2}}
\end{pmatrix}\ ,\ \ \ \ \tilde{\Phi}_a = i\sigma_2\Phi^*_a\ .
\end{align}
According to the non-universal $U(1)$ charges, the Yukawa couplings present in our model are
\begin{align}
-\mathcal{L}_Y &={\bar q'_{iL}}y^{1d}_{ia}\Phi_1 d'_{aR}+{\bar q'_{iL}}y^{1u}_{ia}\tilde{\Phi}_1 u'_{aR} +{\bar l'_{iL}}y^{1e}_{ia}\Phi_1 e'_{aR}+{\bar l'_{iL}}y^{1\nu}_{ia}\tilde{\Phi}_1 \nu'_{aR}\notag\\
&+{\bar q'_{iL}}y^{2d}_{i3}\Phi_2 d'_{3R}+{\bar q'_{iL}}y^{2u}_{i3}\tilde{\Phi}_2 u'_{3R}+{\bar l'_{iL}}y^{2e}_{i3}\Phi_2 e'_{3R} +{\bar l'_{iL}}y^{2\nu}_{i3}\tilde{\Phi}_2 \nu'_{3R} + \text{h.c.}\ .
\end{align}
So, the Yukawa matrices have the following structure
\begin{align}
Y^{1f}=\begin{pmatrix}
y^{1f}_{11} &y^{1f}_{12} &0\\
y^{1f}_{21} &y^{1f}_{22} &0\\
y^{1f}_{31} &y^{1f}_{32} &0
\end{pmatrix}, \ \ \ \ 
Y^{2f}=\begin{pmatrix}
0 &0 &y^{2f}_{13} \\
0 &0 &y^{2f}_{23} \\
0 &0 &y^{2f}_{33} 
\end{pmatrix}\ .\label{eq:yF}  
\end{align}
To analyze the FCNC, it is convenient to rotate the scalar doublets to the Giorgi basis where only one of the CP neutral even components of the doublets acquires VEV while the remaining ones are zero.
Explicitly this corresponds to
\begin{align}
\begin{pmatrix}
\mathcal{H}_1\\
\mathcal{H}_2
\end{pmatrix}=
\begin{pmatrix}
\cos\beta &\sin\beta\\
-\sin\beta &\cos\beta
\end{pmatrix}
\begin{pmatrix}
\Phi_1\\
\Phi_2
\end{pmatrix}\rightarrow 
\mathcal{H}_\alpha =R_{\alpha\beta} \Phi_\beta\ .
\end{align}
In the unitary gauge
\begin{align}
\mathcal{H}_1=\begin{pmatrix}
0\\
\dfrac{v+h}{\sqrt{2}}
\end{pmatrix},\ \ \ \
\mathcal{H}_2=\begin{pmatrix}
\mathcal{H}^+\\
\dfrac{\mathcal{H}^0+i\mathcal{A}^0}{\sqrt{2}}
\end{pmatrix}\ .
\end{align}
The Georgi basis should not be confused with the mass states of the scalar bosons. As discussed in Ref.~\cite{Georgi:1978ri}, in this basis, the $\mathcal{H}_1$ boson gives mass to the SM fermions (the scalar singlet does not couple to SM fermions) and does not generate FCNC.
Therefore, it is not convenient to use the mass eigenstates, a mixture of the scalar singlet and the doublets, when studying the interactions between the scalar sector and the SM fermions. In most observables, the scalar boson is a virtual particle, and the boson that interacts with the SM fermions is the projection onto the subspace formed by the two doublets.
In 2HDM, this feature is very useful since FCNC in the scalar sector can only be generated by $\mathcal{H}_2$. Therefore, in this work, we focus on the CP-even neutral component of this doublet.

In terms of the new basis,
\begin{align}
y^{\alpha f}_{ij}\Phi_\alpha &= x^{\alpha f}_{ij}\mathcal{H}_\alpha\ ,\notag\\  
y^{\alpha f}_{ij}\tilde{\Phi}_\alpha &= x^{\alpha f}_{ij}\tilde{\mathcal{H}}_\alpha\ ,
\end{align}
the rotated yukawa couplings are 
\begin{align}
x^{1f}_{ij}&=\cos\beta\ y^{1f}_{ij} + \sin\beta\ y^{2f}_{ij}\ ,\notag\\
x^{2f}_{ij}&=-\sin\beta\ y^{1f}_{ij} + \cos\beta\ y^{2f}_{ij}\ .\label{eq:x2F} 
\end{align}
Thus
\begin{align*}
\mathcal{L}_Y=&-{\bar  q'_{iL}}x^{\alpha d}_{ij}\mathcal{H}_\alpha d'_{jR}-{\bar q'_{iL}}x^{\alpha u}_{ij}\tilde{\mathcal{H}}_\alpha u'_{jR}-{\bar l'_{iL}}x^{\alpha e}_{ij}\mathcal{H}_\alpha e'_{jR}-{\bar l'_{iL}}x^{\alpha \nu}_{ij}\tilde{\mathcal{H}}_\alpha \nu'_{jR}+ h.c.\ ,\\
=&-{\bar q'_{iL}}x^{1 d}_{ij}\mathcal{H}_1 d'_{jR}-{\bar q'_{iL}}x^{2 d}_{ij}\mathcal{H}_2 d'_{jR}
-{\bar q'_{iL}}x^{1 u}_{ij}\tilde{\mathcal{H}}_1 u'_{jR}-{\bar q'_{iL}}x^{2 u}_{ij}\tilde{\mathcal{H}}_2 u'_{jR}\\
&-{\bar l'_{iL}}x^{1 e}_{ij}\mathcal{H}_1 e'_{jR}-{\bar l'_{iL}}x^{2 e}_{ij}\mathcal{H}_2 e'_{jR}
-{\bar l'_{iL}}x^{1 \nu}_{ij}\tilde{\mathcal{H}}_1 \nu'_{jR}-{\bar l'_{iL}}x^{2 \nu}_{ij}\tilde{\mathcal{H}}_2 \nu'_{jR}+ h.c.\ ,\\
=&-\frac{1}{\sqrt{2}}\left(v+h\right)\left({\bar d'_{iL}}x^{1 d}_{ij}d'_{jR}
+{\bar u'_{iL}}x^{1 u}_{ij} u'_{jR}+{\bar e'_{iL}}x^{1 e}_{ij}e'_{jR}
+{\bar\nu'_{iL}}x^{1 \nu}_{ij}\nu'_{jR}\right)\\
&-\frac{1}{\sqrt{2}}\left(\mathcal{H}^0+i\mathcal{A}^0\right)\left({\bar d'_{iL}}x^{2 d}_{ij} d'_{jR}
+{\bar e'_{iL}}x^{2 e}_{ij} e'_{jR}\right)
-\frac{1}{\sqrt{2}}\left(\mathcal{H}^0-i\mathcal{A}^0\right)\left({\bar u'_{iL}}x^{2 u}_{ij} u'_{jR}
+{\bar\nu'_{iL}}x^{2 \nu}_{ij} \nu'_{jR}\right)\\
&-\mathcal{H}^+\left({\bar u'_{iL}}x^{2d}_{ij} d'_{jR}+{\bar\nu'_{iL}}x^{2 e}_{ij} e'_{jR}\right)
+\mathcal{H}^-\left({\bar d'_{iL}}x^{2 u}_{ij} u'_{jR}+{\bar e'_{iL}}x^{2 \nu}_{ij} \nu'_{jR}\right) + h.c.\ ,
\end{align*}
where we have taken into account that $\tilde{\mathcal{H}}_2=\left(\frac{\mathcal{H}^0-i\mathcal{A}^0}{\sqrt{2}},-\mathcal{H}^-\right)^T$. Next, we define the fermion mass eigenstates as:
\begin{align}
f_L \equiv V^{f\dagger}_L f'_L\ ,\ \ \ f_R \equiv V^{f\dagger}_R f'_R\ , 
\end{align} 
with $V^f_{L(R)}$ are appropriated unitary matrices. In terms of the non-prime fields, we get
\begin{align}
\mathcal{L}_Y=
&-\frac{1}{\sqrt{2}}\left(v+h\right)\left({\bar d_{iL}}z^{1d}_{ij}d_{jR}
+{\bar e_{iL}}z^{1 e}_{ij}e_{jR}+{\bar u_{iL}}z^{1 u}_{ij} u_{jR}
+{\bar\nu_{iL}}z^{1 \nu}_{ij}\nu_{jR}\right)\notag\\
&-\frac{1}{\sqrt{2}}\left(\mathcal{H}^0+i\mathcal{A}^0\right)\left({\bar d_{iL}}z^{2 d}_{ij} d_{jR}
+{\bar e_{iL}}z^{2 e}_{ij} e_{jR}\right)
-\frac{1}{\sqrt{2}}\left(\mathcal{H}^0-i\mathcal{A}^0\right)\left({\bar u_{iL}}z^{2 u}_{ij} u_{jR}
+{\bar\nu_{iL}}z^{2 \nu}_{ij} \nu_{jR}\right)\notag\\
&-\mathcal{H}^+\left({\bar u_{iL}}z^{2ud}_{ij} d_{jR}+{\bar\nu_{iL}}z^{2\nu e}_{ij} e_{jR}\right)
+\mathcal{H}^-\left({\bar d_{iL}}z^{2du}_{ij} u_{jR}+{\bar e_{iL}}z^{2e\nu}_{ij} \nu_{jR}\right) + h.c.\ ,
\end{align}
where
\begin{align}
z^{\alpha f}\equiv V^{f\dagger}_L x^{\alpha f}V^f_R\ ,\ \ \ \ z^{2gf}\equiv V^{g\dagger}_L x^{2f}V^f_R\ .
\end{align}
In this way, from Eqs.~(\ref{eq:x2F}), we get
\begin{align}
z^{1f}=&\cos\beta \left(V^{f\dagger}_L  y^{1f}V^f_R\right) +\sin\beta \left(V^{f\dagger}_L y^{2f}V^f_R\right)\ ,\label{eq:z1F}\\
z^{2f}=&-\sin\beta \left(V^{f\dagger}_L y^{1f}V^f_R\right) +\cos\beta \left(V^{f\dagger}_L y^{2f}V^f_R\right)\ ,\label{eq:z2F}\\
z^{2gf}=&-\sin\beta \left(V^{g\dagger}_L y^{1f}V^f_R\right) +\cos\beta \left(V^{g\dagger}_L y^{2f}V^f_R\right)\ .\label{eq:z2GF}
\end{align}
In the Georgi basis, only the CP-odd component of $\mathcal{H}_1$ acquires VEV, and therefore, the interaction of the SM fermions with this doublet generates the masses of quarks and leptons; consequently,  in the mass eigenstates the matrix $z^{1f}$ must be diagonal, i.e.,
\begin{align}
z^{1f}=\begin{pmatrix}
\frac{\sqrt{2}m^f_1}{v}  &0 &0\\
0 & \frac{\sqrt{2}m^f_2}{v} &0\\
0 &0 & \frac{\sqrt{2}m^f_3}{v} \\
\end{pmatrix},\label{eq:z1F_d}
\end{align}
where $m^f_i$ corresponds to the mass of the fermion $f_i$. 
%In what follows, we will assume that
Because in our model, the right-handed quarks and leptons are singlets under the gauge group
we can define the right-handed fermions as:
$f_{j}''=\left(V^f_R\right)_{ij} f_{j}$
 for all $f$,
such that $V^{f}_R$ completely disappears from the Lagrangian.
This transformation leaves all the terms invariant under the gauge group since the gauge singlets are of the form $\bar{f}_{Ri} f_i$ or $\bar{f}_{Ri} \partial_\mu f_i$  and $V^f_R$ is a global transformation. 
 We are not modifying the  Yukawa interaction terms since we are only redefining them.
 In this way, the coupling of fermions to scalar bosons is 
 \begin{align*}
  z^{\alpha f}\equiv V^{f\dagger}_L x^{\alpha f}V^f_R
\rightarrow V^{f\dagger}_L x^{\alpha f}.
 \end{align*}
A consequence of this result is that if the Yukawa coupling of a scalar boson to one of the right-handed fermions is zero in the interaction space, i.e., $y^{\alpha f}_{ij}=0$ for all $j$, then in mass eigenstates, the corresponding  Yukawa coupling is also identically zero, i.e., $V^{f}_{ik}y^{\alpha f}_{kj} =0$ for all $j$.
That is, if the diagonalization matrix $V^{f}_L$ of the left-handed fermions $f$ is given by
\begin{align}
V^f_L=\begin{pmatrix}
v^f_{11} &v^f_{12} &v^f_{13}\\
v^f_{21} &v^f_{22} &v^f_{23}\\
v^f_{31} &v^f_{32} &v^f_{33}
\end{pmatrix},
\end{align}
the Eq. (\ref{eq:yF}) implies that
\begin{align}
V^{f\dagger}_L y^{1f} =& \begin{pmatrix}
v^*_{11}y^{1f}_{11}+v^*_{21}y^{1f}_{21}+v^*_{31}y^{1f}_{31} &v^*_{11}y^{1f}_{12}+v^*_{21}y^{1f}_{22}+v^*_{31}y^{1f}_{32} &0\\
v^*_{12}y^{1f}_{11}+v^*_{22}y^{1f}_{21}+v^*_{32}y^{1f}_{31} &v^*_{12}y^{1f}_{12}+v^*_{22}y^{1f}_{22}+v^*_{32}y^{1f}_{32} &0\\
v^*_{13}y^{1f}_{11}+v^*_{23}y^{1f}_{21}+v^*_{33}y^{1f}_{31} &v^*_{13}y^{1f}_{12}+v^*_{23}y^{1f}_{22}+v^*_{33}y^{1f}_{32} &0
\end{pmatrix}\ , \notag\\
V^{f\dagger}_L y^{2f} =& \begin{pmatrix}
0 &0 &v^*_{11}y^{2f}_{13}+v^*_{21}y^{2f}_{23}+v^*_{31}y^{2f}_{33}\\
0 &0 &v^*_{12}y^{2f}_{13}+v^*_{22}y^{2f}_{23}+v^*_{32}y^{2f}_{33}\\
0 &0 &v^*_{13}y^{2f}_{13}+v^*_{23}y^{2f}_{23}+v^*_{33}y^{2f}_{33}
\end{pmatrix}.\label{eq:vy}
\end{align}
Since the coupling $z^{1f}$ is diagonal, from the relation $z^{1f}=\cos\beta V^{f\dagger}_L y^{1f}+\sin\beta V^{f\dagger}_L y^{2f}$ and from the fact that if $\left(V^{f}_L y^{1f}\right)_{ij}= 0$ then
$\left(V^{f}_L y^{2f}\right)_{ij}\neq 0$ and the opposite, it must be true that each of the contributions must be diagonal.
From the expressions~\eqref{eq:vy} we have
\begin{align}
z^{1f}_{ij}=& z^{1f}_{i}\delta_{ij}
=\cos\beta \left(V^{f}_L y^{1f}\right)_{ij},
\hspace{0.5cm}\text{for any $i$ and $j=1,2$}\ .\notag\\
z^{1f}_{ij}=& z^{1f}_{i}\delta_{ij}
=\sin\beta \left(V^{f}_L y^{2f}\right)_{ij},
\hspace{0.5cm}\text{for any $i$ and $j=3$}\ .  
\label{eq:z1f}
\end{align}
That is to say,  $\left(V^{f}_L y^{1f}\right)_{ij}$  and $\left(V^{f}_L y^{2f}\right)_{ij}$  are  diagonal matrix with  $\left(V^{f}_L y^{1f}\right)_{33}=0$   and 
$\left(V^{f}_L y^{2f}\right)_{11}=\left(V^{f}_L y^{2f}\right)_{22}=0$.
From these results, the Yukawa couplings 
of $\mathcal{H}_2$ with the physical fermions turn out to be diagonal:
\begin{align*}
z^{2f}=
-\sin\beta \left(V^{f}_L y^{1f}\right)_{ij}+\cos\beta \left(V^{f}_L y^{2f}\right)_{ij}=
-\tan\beta z_i^{1f}\delta_{ij}|_{i,j\in (1,2)}+\cot\beta z_i^{1f}\delta_{i3}\ . 
\end{align*}
In the last step we obtained the expressions for $\left(V^{f}_L y^{1f}\right)_{ij}$ and $\left(V^{f}_L y^{2f}\right)_{ij}$ from Eq.~\eqref{eq:z1f}.
In matrix form this result can be written as 
\begin{align}
z^{2f}=\begin{pmatrix}
-\tan\beta z_1^{1f}&0 &0\\
0 &-\tan\beta z_2^{1f} &0\\
0 &0 &\cot\beta z^{1f}_3
\end{pmatrix}\ .
\end{align}
This result is important because it shows that the coupling of the neutral scalars in the mass eigenstates of the SM fermions is diagonal. Therefore, our model does not present FCNC in the scalar sector.
In most cases, the exact values of the matrix $y^{\alpha f}_{ij}$ are entirely unknown, and what we know are the diagonal couplings, $z_{i}=\frac {\sqrt {2}}{v}m_i^f$, where $m_i^f$ is a diagonal matrix whose elements correspond to the fermion masses in the SM so that the Yukawa couplings will be given by
\begin{align*}
 y^{1f}_{ij}=&V^{f\dagger}_{Lik} 
 \frac{\sqrt{2}}{v\cos\beta}m_k^{1f} \delta_{kj}\ ,
 \text{ where $m_3^{1f}=0$}\ .
\end{align*}
Here $m^{1f}$ is a diagonal matrix 
 with the first two eigenvalues equal to the masses of the particles in the SM and a third element equal to zero.
We have  a similar expression for the second term in \eqref{eq:z1F}   
\begin{align*}
 y^{2f}_{ij}=&V^{f\dagger}_{Lik} 
 \frac{\sqrt{2}}{v\sin\beta}m_k^{2f} \delta_{kj}\ ,
 \text{ where $m_1^{2f}=m_2^{2f}=0$}\ .
\end{align*}
Here $m^{2f}$ is a diagonal matrix 
 with the first two eigenvalues equal to zero and a third element equal to the corresponding mass in the SM.
On the other hand, the Yukawa couplings inducing flavor-changing charged currents can be written as shown below:
\begin{align*}
z^{2ud}&=-\sin\beta\left[V^{u\dagger}_L V^d_L\left(V^{d\dagger}_L y^{1d}V^d_R\right)\right]+\cos\beta\left[V^{u\dagger}_L V^d_L\left(V^{d\dagger}_L y^{2d}V^d_R\right)\right]\ ,\\
z^{2du}&=-\sin\beta\left[V^{d\dagger}_L V^u_L\left(V^{u\dagger}_L y^{1u}V^u_R\right)\right]+\cos\beta\left[V^{d\dagger}_L V^u_L\left(V^{u\dagger}_L y^{2u}V^u_R\right)\right]\ ,\\
z^{2\nu e}&=-\sin\beta\left[V^{\nu\dagger}_L V^e_L\left(V^{e\dagger}_L y^{1e}V^e_R\right)\right]+\cos\beta\left[V^{\nu\dagger}_L V^e_L\left(V^{e\dagger}_L y^{2e}V^e_R\right)\right]\ ,\\
z^{2e\nu}&=-\sin\beta\left[V^{e\dagger}_L V^\nu_L\left(V^{\nu\dagger}_L y^{1\nu}V^\nu_R\right)\right]+\cos\beta\left[V^{e\dagger}_L V^\nu_L\left(V^{\nu\dagger}_L y^{1\nu}V^\nu_R\right)\right]\ .
\end{align*}
Remembering that $V_{\text{CKM}}=V^{u\dagger}_LV^d_L\equiv V$ and $V_{\text{PMNS}}=V^{e\dagger}_LV^\nu_L\equiv U$, it is easy to see from eq. (\ref{eq:z2F}) that
\begin{align}
z^{2ud} =V z^{2d}\ ,\ \ \ z^{2du} =V^\dagger z^{2u}\ ,\ \ \ z^{2\nu e}=U^\dagger z^{2e}\ ,\ \ \ z^{2e\nu}=U z^{2\nu}\ .
\end{align} 

\section{$Z-Z'$ mixing}
\label{sec:mixing}
{
 The $U(1)'$ charge assignments of the Higgs fields $Q_{\Phi}'$ in a specific model, generate the $\theta_{Z-Z'}$  mixing~\cite{Langacker:1991pg,Erler:2009jh} 
 \begin{align}
 \theta_{Z-Z'}
= C\frac{g'}{g_1}\left(\frac{m_Z}{m_{Z'}}\right)^2
= -\frac{\sum_it_{3i}Q_{\Phi_i}^{\prime}v_{\Phi_i}^{ 2}}
{\sum_it_{3i}^2v_i^2}
\frac{g'}{g_1}\left(\frac{m_Z}{m_{Z'}}\right)^2
 \ ,
 \end{align}
 where  $t_{3i}$ is the third component of weak isospin of $\Phi_i$,  $g'$ and $g_1(\sim 0.743)$ are the $Z'$  and $Z$ coupling strength constants, respectively and 
 $m_{Z'}$ and $m_{Z}$ its corresponding masses.

  \begin{align}
 \theta_{Z-Z'}
 =& -2\left(x-z+\frac{v_1^2(y-x)}{v_{SM}^2}\right)
   \frac{1}{g_1}\left(\frac{m_Z}{m_{Z'}}\right)^2\notag\\
 =& -2\left(y-z+\frac{v_2^2(x-y)}{v_{SM}^2}\right)
   \frac{1}{g_1}\left(\frac{m_Z}{m_{Z'}}\right)^2
\sim 
   -\frac{2x}{g_1}\left(\frac{m_Z}{m_{Z'}}\right)^2
 \ ,
 \end{align}
In the last step, we use the approximation 
$v_{SM} \gtrsim v_2^2 \gg v_1^2$ and $x\gg z$.
By imposing the condition $\lvert \theta_{Z-Z'}\rvert< 10^{-3}$ which, roughly speaking, is the upper limit of the $\theta_{Z-Z'}$ mixing for the leptophobic model in reference~\cite{Erler:2009jh}.
We impose this bound for our model since the couplings to the leptons are proportional to the coupling $z$, which, by collider constraints is less than $10^{-2}$ for $Z'$ masses below 2TeV where the $Z-Z'$ mixing constraints are strong. 
Under these considerations we obtain $\frac{2\lvert x\rvert}{g_1}\left(\frac{m_Z}{m_{Z'}}\right)^2<10^{-3}$, which is an almost model-independent result~\cite{Erler:2009jh}. This is equivalent to 
\begin{align}
\lvert x\rvert < \left(\frac{m_{Z'}}{4.68\text{TeV}}\right)^2
\end{align}
These constraints are very restrictive for $Z'$ masses below 2TeV as shown in Figure 1. 
}

\section*{Acknowledges}

R. H. B., E.R., Y.G. and L. M acknowledge additional financial support from Minciencias CD82315 CT ICETEX 2021-1080. This research was partly supported by the ``Vicerrectoría de Investigaciones e Interacción Social~(VIIS) de la
Universidad de Nariño'', project numbers 2686, 2679, 2693, 3130.

\bibliographystyle{unsrt}
\bibliography{references}

\end{document}